\documentclass[twocolumn,twocolappendix]{aastex63}
\usepackage{color}
\usepackage{CJKutf8}
\usepackage{blindtext}
\usepackage{comment}
\usepackage{apjfonts} 
\usepackage{amsmath,amstext}

\begin{document}
\begin{CJK}{UTF8}{bsmi}
\newcommand {\ha}{H$\alpha$}
\newcommand {\hb}{H$\beta$\ }
\newcommand {\hi}{\ion{H}{1}\ }
\newcommand {\nii}{[\ion{N}{2}]\ }
\newcommand {\hii}{\ion{H}{2}\ }
\newcommand {\kms}{km~s$^{-1}$ }
\newcommand {\oiii}{[\ion{O}{3}] }
\newcommand {\sii}{[\ion{S}{2}] }
\newcommand {\hst}{\emph{HST}\ }
\newcommand {\chandra}{\emph{Chandra}\ }

\title{A Multiwavelength Survey of Wolf-Rayet Nebulae in the Large Magellanic Cloud}

\author{Clara Shang Hung （洪宇函）}
\affiliation{Institute of Astronomy and Astrophysics, Academia Sinica, No.1, Sec. 4, Roosevelt Rd., Taipei 10617, Taiwan, R.O.C.} 
\affiliation{Summit Public School: K2, El Cerrito, CA 94530, U.S.A.}

\author[0000-0003-1295-8235]{Po-Sheng Ou （歐柏昇）}
\affiliation{Institute of Astronomy and Astrophysics, Academia Sinica, No.1, Sec. 4, Roosevelt Rd., Taipei 10617, Taiwan, R.O.C.} 
\affiliation{Department of Physics, National Taiwan University, No.1, Sec. 4, Roosevelt Rd.,  Taipei 10617, Taiwan, R.O.C.}

\author[0000-0003-3667-574X]{You-Hua Chu （朱有花）}
\affiliation{Institute of Astronomy and Astrophysics, Academia Sinica, No.1, Sec. 4, Roosevelt Rd., Taipei 10617, Taiwan, R.O.C.} 
\affiliation{Department of Physics, National Taiwan University, No.1, Sec. 4, Roosevelt Rd.,  Taipei 10617, Taiwan, R.O.C.} 
\affiliation{Department of Astronomy, University of Illinois, 1002 West Green Street, Urbana, IL 61801, U.S.A. } 
 
\author[0000-0002-4588-6517]{Robert A.\ Gruendl}
\affiliation{Department of Astronomy, University of Illinois, 1002 West Green Street, Urbana, IL 61801, U.S.A. } 

\author[0000-0003-1449-7284]{Chuan-Jui Li （李傳睿）}
\affiliation{Institute of Astronomy and Astrophysics, Academia Sinica, No.1, Sec. 4, Roosevelt Rd., Taipei 10617, Taiwan, R.O.C.}

%

\begin{abstract}
Surveys of Wolf-Rayet (WR) stars in the Large Magellanic Cloud (LMC) have yielded a fairly complete catalog of 154 known stars. We have conducted a comprehensive, multiwavelength study of the interstellar/circumstellar environments of WR stars, using the Magellanic Cloud Emission Line Survey (MCELS) images in the \ha, \oiii, and \sii lines; \emph{Spitzer Space Telescope} 8 and 24 $\mu$m images; Blanco 4m Telescope \ha\ CCD images; and Australian Telescope Compact Array (ATCA) + Parkes Telescope \hi data cube of the LMC. We have also examined whether the WR stars are in OB associations, classified the \hii environments of WR stars, and used this information to qualitatively assess the WR stars' evolutionary stages. The 30 Dor giant \hii region has active star formation and hosts young massive clusters, thus we have made statistical analyses for 30 Dor and the rest of the LMC both separately and altogether.  Due to the presence of massive young clusters, the WR population in 30 Dor is quite different from that from elsewhere in the LMC.  We find small bubbles ($<$50 pc diameter) around $\sim$12\% of WR stars in the LMC, most of which are WN stars and not in OB associations.  The scarcity of small WR bubbles is discussed.  Spectroscopic analyses of abundances are needed to determine whether the small WR bubbles contain interstellar medium or circumstellar medium.  Implications of the statistics of interstellar environments and OB associations around WR stars are discussed.  Multiwavelength images of each LMC WR star are presented.
\end{abstract}

\keywords{ISM: bubbles-- stars: Wolf–Rayet}

\section{Introduction}  \label{sec:intro}
Wolf-Rayet (WR) stars are often surrounded by beautiful ring-shaped nebulae in \ha\ images. Since the discovery of the first three WR ring nebulae, NGC\,2359, NGC\,6888, and S308 \citep{Johnson1965}, systematic surveys of WR nebulae have been conducted in our Galaxy \citep[e.g.,][]{Chu1981,Heckathorn1982, CTK1983, Miller1993, Marston1994a, Marston1994b, Stock2010} and the Large Magellanic Cloud \citep[LMC; e.g.,][]{Chu1980,Dopita1994}. It is found that ring nebulae around WR stars can have a variety of morphologies and sizes, possibly indicating different formation mechanisms and evolutionary stages. To carry out a comprehensive investigation of ring nebulae around WR stars, it is beneficial to compile a complete inventory of WR stars in a galaxy and use deep nebular line images to search for associated nebulae. 

It is difficult to achieve completeness in the catalog of WR stars in our Galaxy because of the heavy extinction along the Galactic plane. Even the distances of the Galactic WR stars were uncertain before the Gaia data became available \citep{Gaia2018}. It would thus be difficult to perform statistical and quantitative analyses on the WR nebulae in our Galaxy. The LMC, on the other hand, is at a known 50 kpc distance \citep{Piet2019} with little internal and foreground extinction; thus, it is possible to achieve a more complete inventory of WR stars and carry out a thorough multiwavelength search for associated nebulae. 

An initial search of WR stars in the LMC by \citet{Westerlund1964} found 58 objects, and subsequent searches have increased the number of known LMC WR stars to 80 by \citet{Fehrenbach1976}; 100 by \citet{Breysacher1981}; 135 by \citet{Breysacher1999}; and more recently 154 by \citet{Neugent2018}. The latest LMC WR star catalog is likely fairly complete. 

The search for WR nebulae in the LMC began with \citet{Chu1980} using photographic plates taken with the 0.61 m Curtis Schmidt Telescope. Later, \citet{Dopita1994} used the Australian National University 2.3 m Telescope with a spectrograph in the imaging mode, as well as an imager, to conduct a deeper survey on WR nebulae. \citet{Stock2010} used digitized \ha\ photographic plates taken with the 1.2 m UK Schmidt Telescope to search for WR nebulae in the LMC. The former two surveys used \ha\ and \oiii images, while the latter used only \ha\ images. Meanwhile, the Magellanic Cloud Emission Line Survey (MCELS) has been conducted with the Curtis Schmidt Telescope and a CCD camera, providing \ha, [\ion{O}{3}], and \sii images for the entire LMC \citep{Smith1999}; furthermore, the \emph{Spitzer Space Telescope} has surveyed the LMC \citep{Meixner2006} in infrared passbands, and the Australian Telescope Compact Array (ATCA) has surveyed the \hi in the LMC with high resolution \citep{Kim2003}. Using these new surveys of neutral and ionized gas and the nearly complete catalog of WR stars in the LMC, we are able to, for the first time, probe the multiphase interstellar environment of WR stars and investigate interactions between the stars and their ambient medium.

This paper reports our survey and analysis of WR nebulae in the LMC. Section 2 expands upon the expected coevolution of a WR star and its environment, Section 3 describes the data and methodology used to conduct this investigation, Section 4 presents the results, Section 5 discusses the implications of our results, and Section 6 summarizes this work.

\section{Coevolution of WR Stars and Their Ambient Medium} 
When WR nebulae were first discovered, neither the evolutionary history of WR stars nor the formation mechanisms of the nebulae were clearly understood. Over the years, stellar evolution models have shown that WR stars are just evolved massive stars, of which some are well beyond the main sequence (MS) and some are still burning hydrogen \citep{Langer2012}. The massive stars' winds during different evolutionary stages have become better known. Hydrodynamic simulations of interactions between massive stars' winds and their ambient medium have been made \citep[e.g.,][]{Garcia1996II, Garcia1996I}, and their results can be used to guide our search for WR nebulae. In this section, we first summarize our early perception of WR nebulae based on data with inadequate resolutions and our incomplete knowledge of stellar evolution and stellar wind interaction. The misunderstandings should be superseded by our current understanding based on numerical simulations made with improved knowledge of stellar evolution and its associated stellar wind properties. 
\subsection{Previous Perception} 
WR stars are massive stars characterized by broad emission lines in their spectra, indicating the presence of fast stellar winds. Among massive stars, WR stars' winds have the highest mechanical luminosities. The powerful WR winds sweep up the ambient medium to form shell structures that appear as ``arcs'' or ``rings'' in \ha\ images and are called ``ring nebulae.''

WR nebula surveys used only morphological information in \ha\ images to identify ring nebulae. More often than not, a WR star is projected near a curved nebular filament, and the physical relationship between the WR star and the nebular feature is tantalizing but not confidently established. Despite the uncertainty, such objects were still identified as WR ring nebulae. Sometimes, a round \hii\ region with a WR star in its central region is also identified as a ring nebula.  Thus, WR nebulae became a heterogeneous class of objects.

To investigate the nature of WR ring nebulae, \citet{Chu1981} incorporated internal kinematic properties of the WR nebulae and classified WR ring nebulae into three categories, based on the kinematics and morphologies of the nebulae: \begin{itemize}

\item{} W-type nebulae are wind-blown bubbles whose dynamic ages are smaller than the lifetime of a WR phase to ensure that the WR winds are responsible for the formation of the bubbles.  Such a ring nebula shows a fine filamentary shell around the central star, which is usually projected near the center or toward the brightest part of the nebula.

\item{} E-type nebulae consist of stellar ejecta. The E-type nebulae were introduced to have chaotic and irregular expansions and clumpy morphologies. However, it was later realized that the chaotic expansion reported for M1-67 and RCW\,58 \citep{ChuTreffers1981, Chu1982a} was an artifact caused by the large entrance apertures of the Fabry-Perot scanning observations, whereas long-slit high-dispersion spectra actually show uniform expansion pattern despite the large density fluctuations \citep{Solf1982, Chu1988}. Chemical abundance observations of the E-type WR nebulae show enrichment of nitrogen \citep[e.g.,][]{Esteban1991}, confirming that they consist of ejected stellar material, which is called circumstellar medium (CSM).

\item{} R-type nebulae are characterized as ``radiatively excited \hii regions," consisting of interstellar medium (ISM).  $R_{\rm a}$ denotes amorphous \hii\ regions, while $R_{\rm s}$ denotes shell nebulae whose dynamic ages are much larger than the lifetime of the WR phase.

\end{itemize}

When deeper and higher-resolution images are available, it becomes clear that the above classification is too primitive. A more sophisticated approach is needed. The perception of WR nebulae in this subsection should be superseded by the understanding outlined in the next subsection.

\subsection{Current Understanding} 

The current knowledge of WR nebulae derives from a better understanding of stellar evolution. We now know that WR stars evolve from MS O stars through either the luminous blue variable (LBV) or red supergiant (RSG) phase \citep{Langer2012}. \citet{Garcia1996II, Garcia1996I}  showed for the first time a hydrodynamic model of the gaseous environment coevolving with the central star from the MS phase to the WR phase. Similar hydrodynamic calculations have been carried out by \citet{Dwarkadas2007}, \citet{Toala2011}, and \citet{vanMarle2012}, all of which have produced qualitatively similar results, as described below. 
 
The WR stars that evolve from the MS through the LBV phase are the most massive, e.g., $\gtrsim$ 60 $M_\odot$, while those that evolve from MS via the RSG phase are less massive, e.g., $\sim$ 35 $M_\odot$. In the MS phase, massive stars lose mass at rates of $\sim$ 10$^{-6}$ $M_\odot$ yr$^{-1}$ in the form of fast stellar winds with terminal velocities of $\sim$2000--3000 km s$^{-1}$ \citep{Prinja1990, Puls1996}. During this stage, the MS stellar wind sweeps up the ambient ISM to form an interstellar bubble \citep{Weaver1977}. As a massive star evolves off the MS and into the LBV or RSG phase, the mass loss takes place in the form of copious slow winds, of which the wind velocities are $\sim$10--50 km s$^{-1}$ and mass-loss rates $\sim$ 10$^{-4}$ $M_\odot$ yr$^{-1}$ \citep{vanLoon2005}. The stellar material lost in this wind expands slowly away from the star, forming a small circumstellar nebula (consisting of CSM) inside the cavity of the interstellar bubble formed previously when the star was in the MS stage.

When the star enters a WR phase, the stellar wind velocity and mass-loss rate increase considerably, with terminal velocities of $\sim$1000--3000 km s$^{-1}$ and mass-loss rates of $\sim$ 10$^{-5}$ $M_\odot$ yr$^{-1}$ \citep{Nugis2000}. This fast WR wind sweeps up the slowly expanding CSM into a shell, called a circumstellar bubble. Therefore, we expect to see a nested shell structure consisting of a small circumstellar bubble within a larger interstellar bubble around the WR star.

If a WR star has a high velocity with respect to the ambient ISM, either ejected from a cluster or kicked by the supernova explosion of a binary companion, it is a runaway star. For a WR star moving in the ISM, an interstellar bow shock will form in the direction of the star's trajectory. However, if a WR star is surrounded by CSM, which travels together with the WR star, a circumstellar bubble can still form, but the leading edge will be compressed by the ISM and become sharper, denser, and brighter, as seen in the circumstellar bubble around the LBV star S119 \citep{Danforth2001}. 

We have also come to understand that whether a WR star is an isolated single star or located in a cluster environment as a member of an OB association affects the evolution of shell nebulae around the WR star. In the presence of other massive stars, the stellar winds and supernova explosions from all stars collectively form a large shell, called a superbubble \citep{McCray1987}. The formation of a circumstellar WR bubble inside a superbubble is trickier than inside a small interstellar bubble because of possible impacts of stellar winds and supernova explosions from neighboring massive stars.

The largest interstellar structures in a galaxy are supergiant shells (SGSs) whose sizes can approach 1000 pc. The formation of such large shells require multiple generations of star formation to provide the energy. If a WR star is in the low-density interior of an SGS, it may not have a detectable interstellar bubble, although its circumstellar bubble should still be present.

With the known characteristics of the different stages in the evolution of WR stars, we expect to see circumstellar bubbles, interstellar bubbles, superbubbles, and SGSs around WR stars. However, to discern between circumstellar bubbles and interstellar bubbles, observations of nebular abundances are needed, but they are not available to us. Thus, we can only use morphological information to search for small bubbles centered on individual WR stars, superbubbles around WR stars and their host OB associations, and SGSs on the largest scales.

\section{Data and Methodology} 
We have used multiwavelength images to examine the ionized and neutral gaseous environments of the LMC WR stars. We used optical emission-line images to examine the distribution and excitation of ionized gas, the \hi 21 cm line data cube to investigate the distribution and kinematics of neutral atomic gas, 8 $\mu$m images to identify the partially dissociated regions, and 24 $\mu$m images to locate emission from heated dust.
\subsection{Data Used} 

MCELS1 is an emission-line survey of the Magellanic Clouds \citep{Smith1999}. The survey used the Curtis Schmidt Telescope at Cerro Tololo Inter-American Observatory (CTIO) and a CCD camera to take images in the \ha, [\ion{O}{3}] $\lambda$5007, and [\ion{S}{2}] $\lambda\lambda$6716, 6731 bands, as well as two continuum bands centered at $\lambda$6850 and $\lambda$5130. In this paper, we only used the emission-line images without continuum subtraction. The \ha\ images show the overall distribution of ionized gas; the \oiii images, compared with the \ha\ images, reveal the excitation of ionized gas; and the \sii images diagnose shocks and ionization fronts. 

A higher-resolution \ha\ survey of the Magellanic Clouds has been conducted with the MOSAIC II camera on the Blanco 4m Telescope at CTIO; this survey has been called MCELS2 (PI: You-Hua Chu). The MOSAIC II camera, which is a mosaic of eight SITe 4096 $\times$ 2048 CCDs, has a pixel size of 0\farcs27 $\times$ 0\farcs27 and a field-of-view of 36$'$ $\times$ 36$'$.  An \ha\ filter with 80 \AA\ width was used to record images. In celebration of the 30-year anniversary of the Hubble Space Telescope (HST), a public release of HST H$\alpha$ image of NGC 2020 became available, and this image is used for WR 71.

The \hi data cube was constructed by \citet{Kim2003} using the ATCA survey data in conjunction with single-dish observations from the Parkes Telescope. This data cube covers 11.1$^{\circ} \times 12.4^{\circ}$ of the sky with an angular resolution of 1\farcm0 and a pixel size of 20$''$. The heliocentric velocity coverage, $-$33 to +627 km s$^{-1}$, is centered at $\sim$300 km s$^{-1}$, the bulk radial velocity of the LMC. We use the zeroth moment map to present the \hi column densities and position-velocity plots to show the kinematic structures of \hi gas.

The \emph{Spitzer Space Telescope} Infrared Array Camera \citep[IRAC;][]{Fazio2004} 8 $\mu$m images are used to identify polycyclic aromatic hydrocarbons (PAHs) associated with partially dissociated regions, and the Multiband Imaging Photometer for Spitzer \citep[MIPS;][]{Rieke2004}
24 $\mu$m images to reveal emission from heated dust.  The \emph{Spitzer} images of the LMC were from the Legacy Program Surveying the Agents of a Galaxy's Evolution \citep[SAGE;][]{Meixner2006}.

\subsection{Methodology} 
We have used our understanding of the coevolution of WR stars in their ambient medium, as detailed in Section 2.2, to guide our search for nebular shells around WR stars. More specifically, to identify wind-blown bubbles, we search for small shells less than $\sim$50 pc in diameter with the WR star in a preferred location indicating that wind-ISM/CSM interaction is responsible for forming the shell. For example, the WR star should be either near the center of a uniform shell or closer to the brighter part of a nonuniform shell; if the structure is a bow-shock-like arc, then the WR star should be near the arc's center of curvature. We identify superbubbles by their large shell sizes, much larger than 50 pc in diameter, or by a central cluster or OB association. For example, the shell around the R136 cluster measures 37.5 pc $\times$ 21 pc in size yet we denote it as a superbubble because of its prominent central cluster. On still a larger scale, we identify SGSs that are greater than $\sim$500 pc in diameter to contextualize the extended star formation history in the WR star's large-scale environment.

To search for nebular shells around WR stars, we first examined the MCELS1 \ha\ images to obtain a census of ionized gas around the WR stars. We noted the nebular morphology, specifically filaments, arcs, and shell structure in the vicinity of the star. We measured and recorded the dimensions of shell-like features centered on or surrounding the star. In cases where there were no shells, we also noted any emission nebulae around the star. We further used MCELS2 \ha\ images for a high-resolution view of the filamentary structures in the WR nebula.

To corroborate our findings on the WR nebulae from the \ha\ images, we examined MCELS1 \oiii and \sii images to gain insight into the nebular excitation and physical conditions. For early WR stars, i.e., high stellar effective temperatures, we expect to see high \oiii$\lambda$5007/\ha\ ratios; thus, bubbles of stars with early spectral types stand out against the ambient medium in the \oiii images better than in the \ha\ images. Conversely, for late WR stars, the \oiii$\lambda$5007/\ha\ ratios are low, and the nebulae are usually not detected in \oiii images.

Beyond examining optical images of ionized gas around WR stars, we also extracted the \hi zeroth moment map and position-velocity plots for north-south and east-west strips of 20$''$ width in order to show the \hi column density distribution and the kinematics of the \hi gas, respectively. A low-intensity region in the zeroth moment map indicates a depression in the \hi column density. A bow-shaped velocity split in a position-velocity plot implies an expanding shell structure.

The dust emissions highlighted by \emph{Spitzer}'s IRAC 8 $\mu$m and MIPS 24 $\mu$m images were used to gain a more comprehensive understanding of the physical conditions of the WR star's interstellar and circumstellar environments.  Some very late-type WR stars with dusty envelopes are also very bright in the 24 $\mu$m band.

To see whether a WR star is a member of an OB association \citep{LH1970}, we compared the WR star's location with the boundaries of OB associations provided in the finding charts in P.\ Lucke's Ph.D. thesis. WR stars not within the boundaries of an OB association but are within 100 pc to an OB association are also noted, with single parentheses for within 50 pc and double parentheses for 50--100 pc from the center of the OB association.  Our results are more quantitative and complete than the compilation by \citet{Neugent2018}, who have missed a few associations, such as WR 38 and 39 in LH45, WR 43 in LH 50, WR 53 in LH 62, and WR 55 in LH 61. We record this information in case the WR star is a runaway star from the OB association.  

We also identify the \hii regions around the WR stars in the \citet{DEM1976} and \citet{Henize1956} \hii region catalogs. We further classify their \hii regions in three morphological classes. Class 1 refers to a bright, amorphous \hii region without filamentary or shell structures. Class 2 indicates that the \hii region has a large shell structure, e.g., a superbubble, and it is further divided into two subclasses, with 2a referring to a highly nonuniform shell with the ionizing stars/cluster closer to the brightest rim of the shell and 2b referring to a relatively uniform shell around the ionizing stars/cluster. Lastly, Class 3 indicates a very low-density environment of the WR star. It is further divided into three subclasses: with 3a indicating faint but detectable diffuse emission in the vicinity of the star, 3b indicating the low-density interior of a large shell, and 3c reflecting no detectable diffuse emission in a large-scale diffuse field.

\section{Results} 
\begin{figure*}[tbh]
\centering
\includegraphics[width=\textwidth]{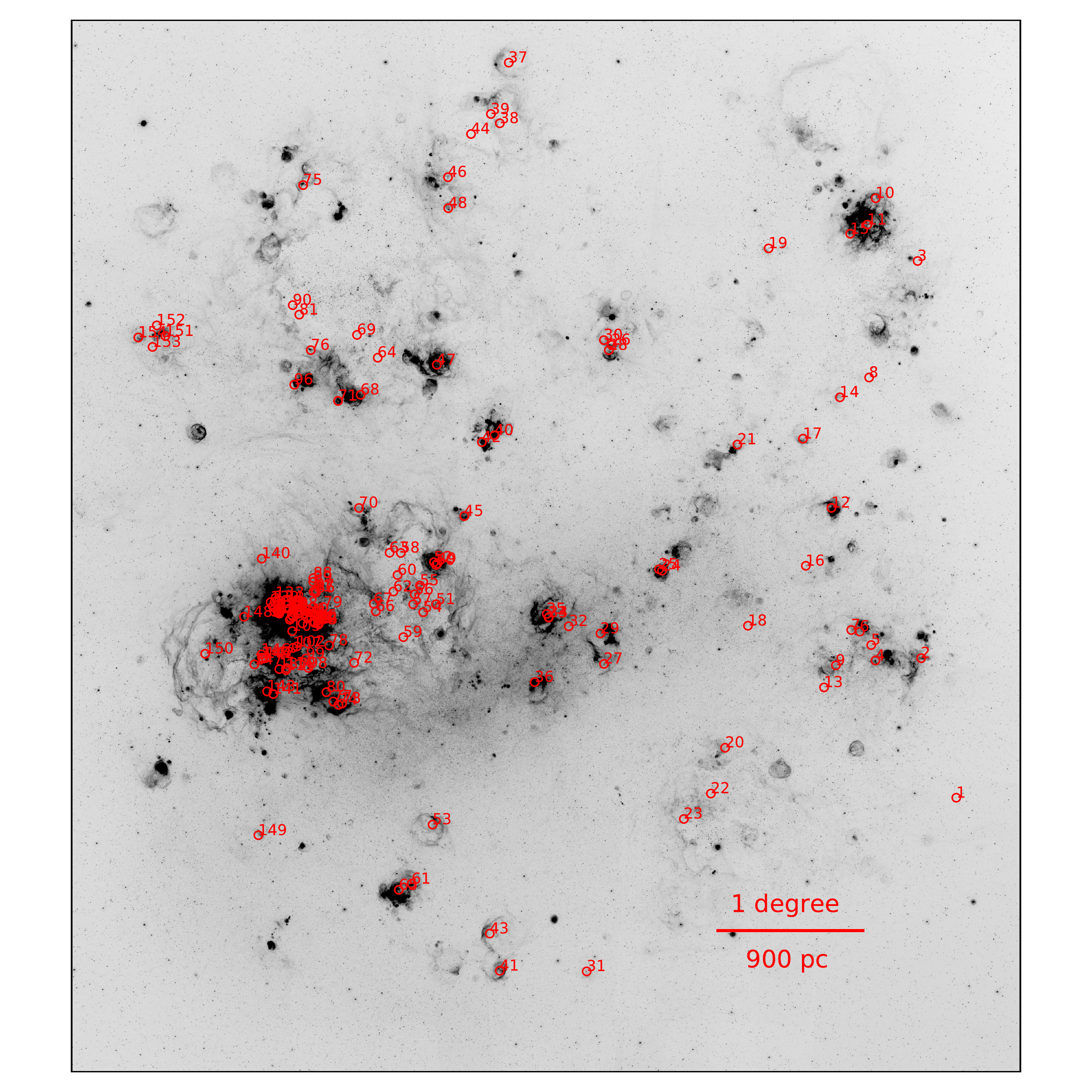}
\caption{The positions of all the 154 WR stars in the LMC. The running numbers are from \citet{Neugent2018}.}
\label{fig:map}
\end{figure*}
\figsetstart
\figsetnum{2}
\figsettitle{LMC WR stars}

\figsetgrpstart
\figsetgrpnum{2.62}
\figsetgrptitle{WR71}
\figsetplot{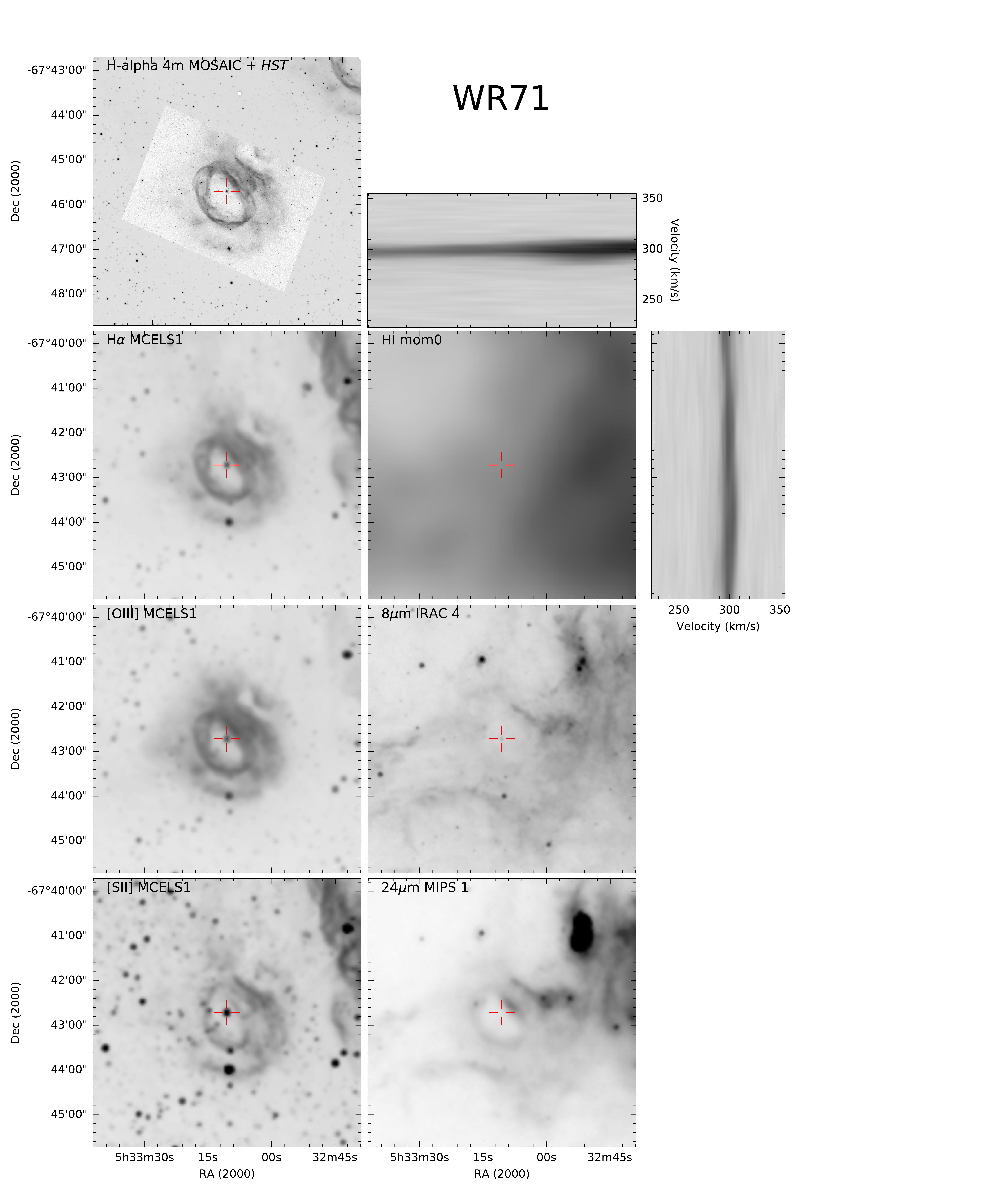}
\figsetgrpnote{Multiwavelength images of WR71.  The WR star is marked by a red cross in
the images.  The origin and passband information is marked on the upper 
left corner of each image.  The HI position-velocity plots along the EW 
and NS direction centered on the WR star are plotted above and to the 
right of the 0th moment map, respectively. }
\figsetgrpend

\figsetend
\begin{figure*}
\plotone{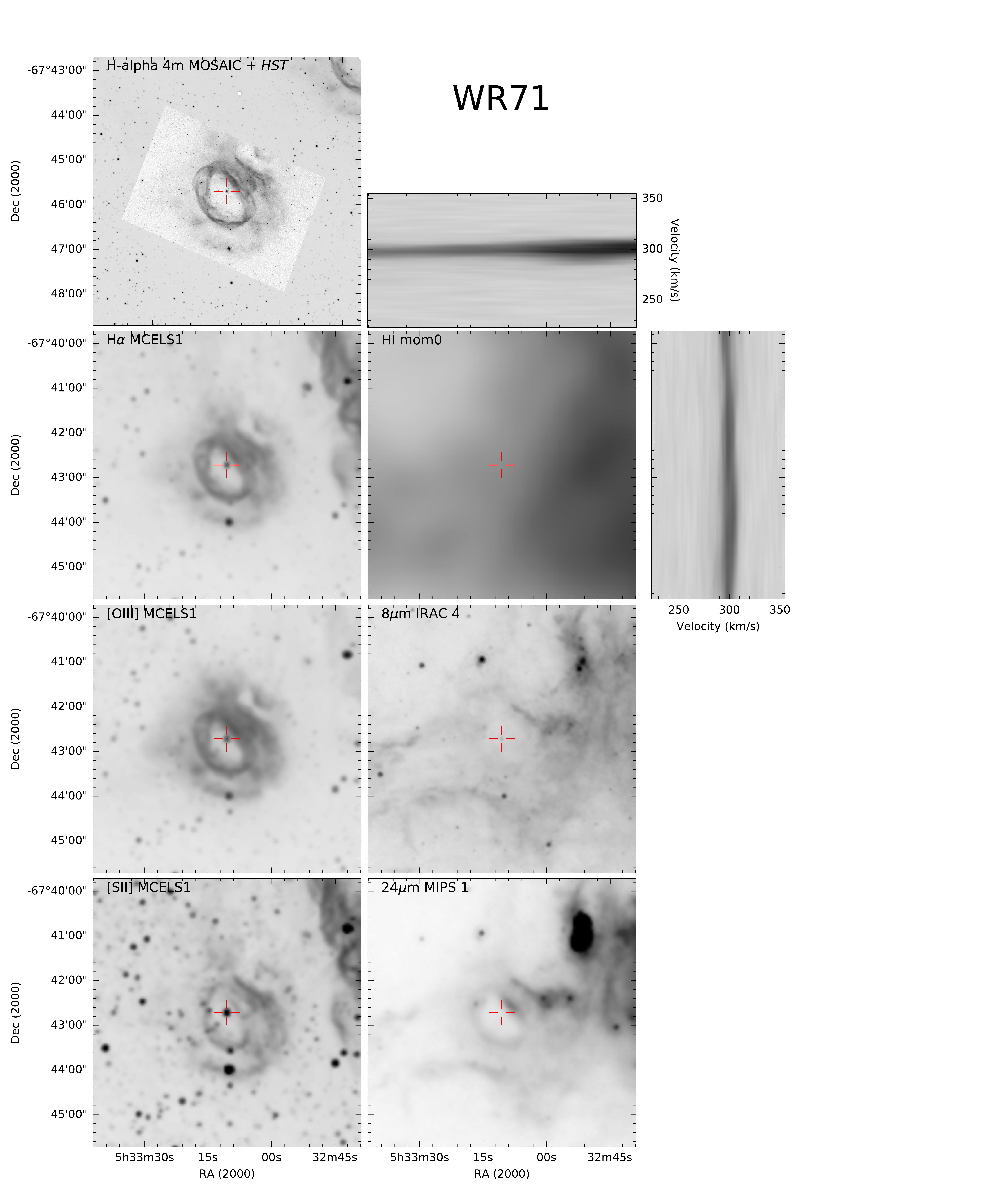}
\caption{Multiwavelength images of WR stars in the LMC. 
Each figure has the WR star name (or names) labeled on the top
and position marked by a red cross or circle.  The origin and passband information is marked on the upper-left corner of each panel. The H I position-velocity plots along the EW and NS directions are plotted above and to the right of the zeroth moment map respectively. The figure of WR 71 is shown here as an example. The complete figure set (99 figures) is available in the online journal.}
\label{fig:allWR}
\end{figure*}

The 154 WR stars from \citet{Neugent2018} are marked on the MCELS1 H$\alpha$ image of the LMC in Figure~\ref{fig:map}. 
For each WR star or a close group of WR stars, we make a figure that includes the MCELS2 H$\alpha$ image; MCELS1 \ha, [\ion{O}{3}], and \sii images; \emph{Spitzer} 8 and 24 $\mu$m images; and \hi zeroth moment map and position-velocity plots.  An example is shown in Figure~\ref{fig:allWR}, and the rest in Fig. Set 2, LMC WR stars.
The complete descriptions of individual WR stars and their environments are provided in the Appendix.  

We summarize our findings on the stellar and interstellar environments of all 154 WR stars in the LMC in Table 1. Column 1 is the running number of the WR star in \citet{Neugent2018}; columns 2 and 3 are the R.A. and decl. of the WR star; columns 4 and 5 are the running numbers of the WR star from \citet{Breysacher1999} and \citet{Breysacher1981}, respectively; column 6 is the spectral type of the WR star from \citet{Neugent2018}; and column 7 is the host OB association from \citet{LH1970}, where single parentheses mark WR stars not in the OB association but are within 50 pc and double parentheses denote WR stars beyond 50 pc but within 100 pc to the center. Columns 8 and 9 are the associated \hii regions from the \citet{Henize1956} and \citet{DEM1976} catalogs; column 10 is the \hii morphology classification; columns 11 and 12 are the dimensions of any bubble and superbubble associated with the WR star, respectively; column 13 lists the \hii supergiant shells from \citet{Meaburn1980}; and column 14 lists the \hi giant and supergiant shells from \citet{Kim1999}. Note that the use of single parentheses around the \hii morphology and bubble/superbubble sizes indicate uncertainty. 

In the statistical analysis of WR stars and their associated nebular environments, the 30 Doradus (30 Dor) region needs special consideration because of its extreme properties. 30 Dor is an archetypical giant \hii region where star formation is characterized as ``starburst'' and internal gas dynamics is violent.  Early low-resolution radio images of 30 Dor show three peaks in 30 Dor \citep{LeMarne1968}: the brightest peak, 30 Dor A (N157A), corresponds to roughly the 100 pc radius region centered on the R136 cluster, or OB association LH 100; the second brightest peak, 30 Dor B (N157B), corresponds to the \hii region around LH 99 and contains a supernova remnant with a pulsar wind nebula \citep{Wang2001}; and the third peak, 30 Dor C (N157C), corresponds to the superbubble around LH 90.  These three components are encompassed in DEM\,L263 \citep{DEM1976} or N157 \citep{Henize1956}, and within this region exist 43 known WR stars, almost 30\% of the whole WR population in the LMC.  In the following statistical analyses, we will treat 30 Dor and the rest of the LMC separately, as well as together for the entire LMC. These results will be compared and discussed in the next section.

Of all the variously sized shells encompassing the WR stars, small bubbles are the most relevant to each individual star as the WR wind may be directly responsible for forming the bubble. Thus, we have compiled all small bubbles from Table 1 into Table 2 to examine the statistics of the bubble sizes and WR spectral types. The small bubble around WR 97 in 30 Dor is listed in Table 2 but separated by a horizontal line at the bottom. The H$\alpha$ images of the 18 small bubbles are shown in Figure~\ref{fig:bubble}. 
We have compiled all superbubbles outside and inside 30 Dor from Table 1 into Tables 3 and 4, respectively, to probe the relationship between WR stars, OB associations, and superbubbles. Images of example superbubbles are shown in Figure~\ref{fig:superbubble}.

As a reference for the tendency of WR stars' locations in OB associations, we have compiled in Table 5 numbers of WR stars for different spectral types and for the entire LMC, the LMC excluding 30 Dor, and 30 Dor, respectively. To obtain the statistical significance, we combine spectral types into subgroups: WN2-4 (early WN), WN5-6 (mid WN), WN7-L (late WN), WC4, WC5-6, WO3-4, and ``other'' for the small number of remaining objects. The LMC WR star catalog of \citet{Neugent2018} has 154 entries, but two entries contain double WR stars and thus the total number of stars in Table 5 is 156.  The percentage of WR stars that are in LH OB associations is given in parentheses after the numbers of WR stars in OB associations.  It is quite clear that the WR population in the LMC is dominated by WN stars, $\sim$80\%, inside or ouside 30 Dor alike.  Within the WN population, there exist more early-type than late-type objects for the entire LMC; however, the WN population differs significantly inside and outside 30 Dor.  While outside 30 Dor 73\% of WN stars are of WN2-4 types and 9\% are WN5-6, inside 30 Dor, only 19\% of WN stars are WN2-4 and 62\% are WN5-6. Overall, $\sim$50\% of the WR stars are in OB associations; however, WN5-6 and WC stars are more likely to be in OB associations than WN2-4 and WN7-L.  Outside 30 Dor, WC stars have the highest percentage to be in OB associations, while inside 30 Dor, WN5-6 stars have the highest percentage to be in OB associations.


Table 6 presents the statistics of the WR star population in the LMC and the fraction of each spectral type with bubbles/superbubbles.  In this table, column 1 gives the spectral type grouping, column 2 presents the number (and percentage) of WR stars with bubbles, column 3 presents the number of WR stars with bubbles in OB associations, column 4 presents the number (and percentage) of WR stars in superbubbles, and column 5 presents the number of WR stars with superbubbles in OB associations for all WR stars in the LMC; columns 6-9 present the same information as columns 2-5 for all WR stars in the LMC excluding 30 Dor, respectively; column 10 presents the number (and percentage) of WR stars in superbubbles in 30 Dor and column 11 presents the number of 30 Dor WR stars in superbubbles in OB associations.

In 30 Dor, few small bubbles are identified: WR 97 is surrounded by a small bubble-like structure, while WR 118/119/120 are projected on the southern rim of a triangular shell structure that could be connected to the central superbubble around the R136 cluster, and thus, the bubble nature of the triangular shell is highly uncertain.  Several large shell structures in 30 Dor can be identified as superbubbles \citep{Chu1994}, and WR stars projected within superbubbles in 30 Dor are compiled in Table 4, along with the dimensions of the superbubble and host OB associations. Owing to the scarcity of small WR bubbles in 30 Dor, we only give statistics of WR stars in superbubbles in 30 Dor in Table 6.

We define the \hii morphology classes to represent the evolutionary stages of the ISM surrounding the WR stars.  Thus, the correlation between the spectral types and the \hii morphology classes may be used to diagnose the WR star's progression in its evolution. We have tallied the numbers of WR stars for different spectral types in \ion{H}{2} regions of different morphological classes in Table 7.  The results are reported separately for 30 Dor and the LMC excluding 30 Dor, as well as the entire LMC.

\begin{figure*}[tbh]
\centering
\includegraphics[width=\textwidth]{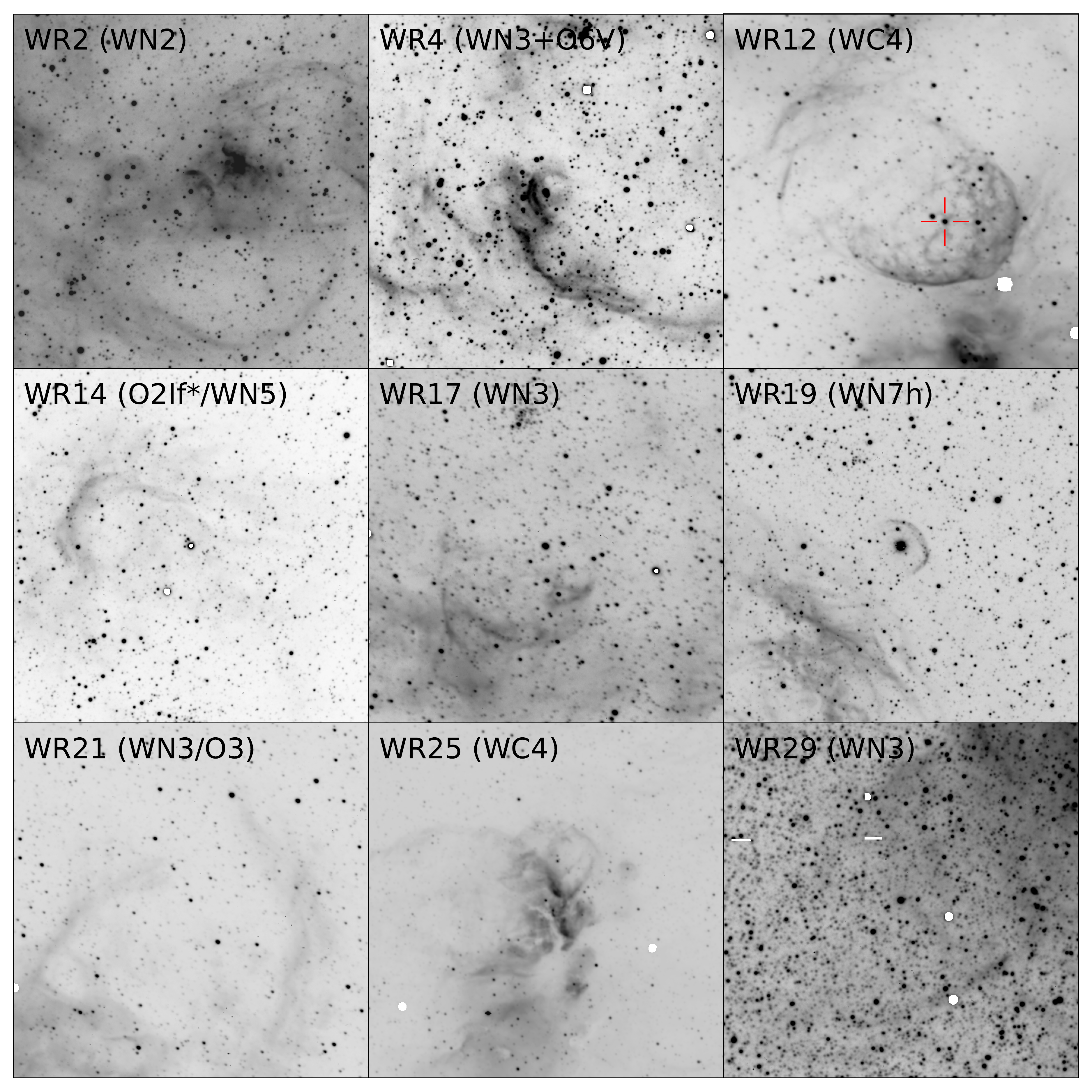}
\caption{4m MOSAIC H$\alpha$ images of small bubbles around WR stars in the LMC. The field of view of each panel is 4'$\times$4'. The WR star is at the center of the panel except WR 12, which is marked by a red cross.}
\label{fig:bubble}
\end{figure*}
\begin{figure*}[tbh]
\centering
\figurenum{3}
\includegraphics[width=\textwidth]{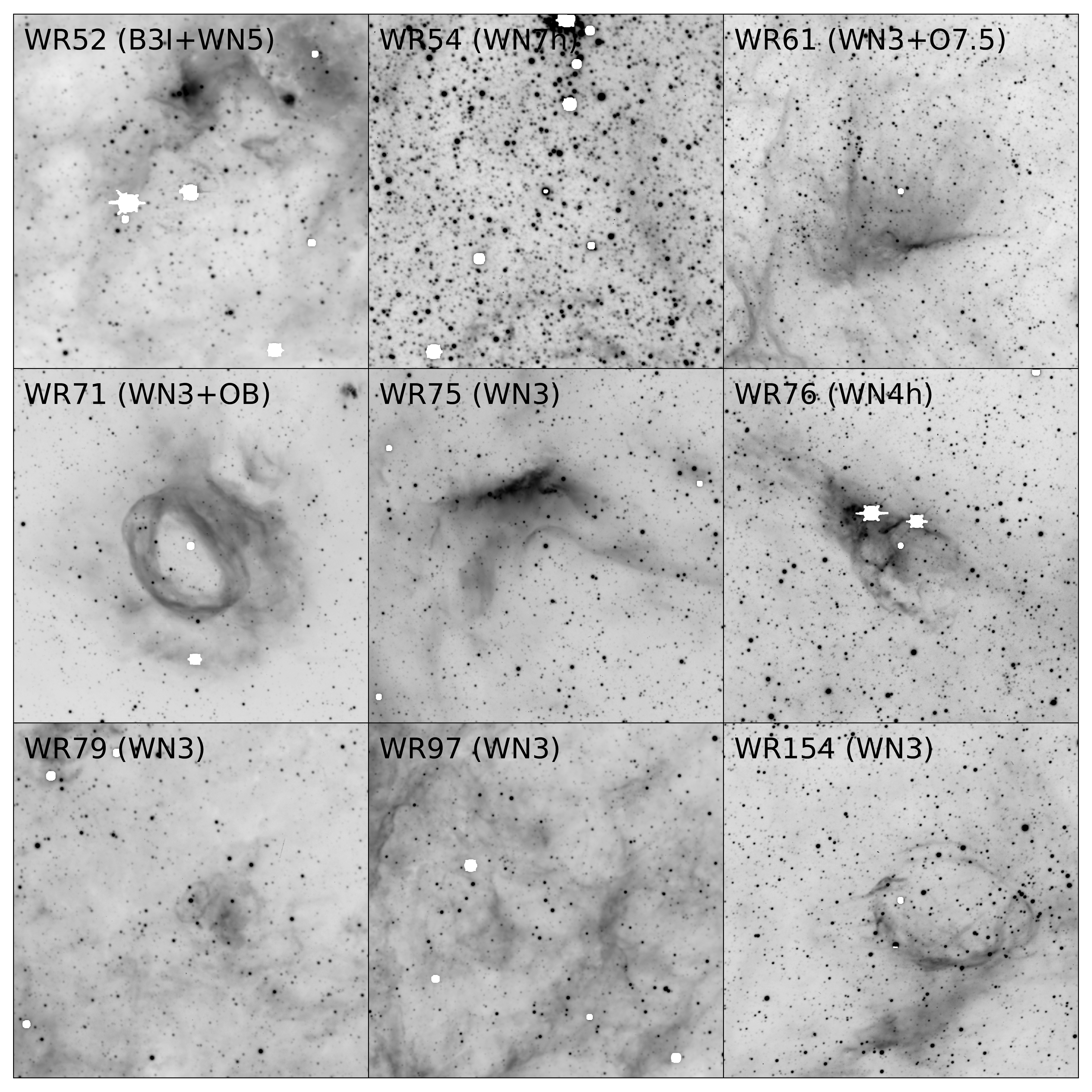}
\caption{(cont.) 4m MOSAIC H$\alpha$ images of small bubbles around WR stars in the LMC. The field of view of each panel is 4'$\times$4'. The WR star is at the center of the panel except WR12, which is marked by a red cross.}
\end{figure*}
\begin{figure*}[tbh]
\centering
\includegraphics[width=15cm]{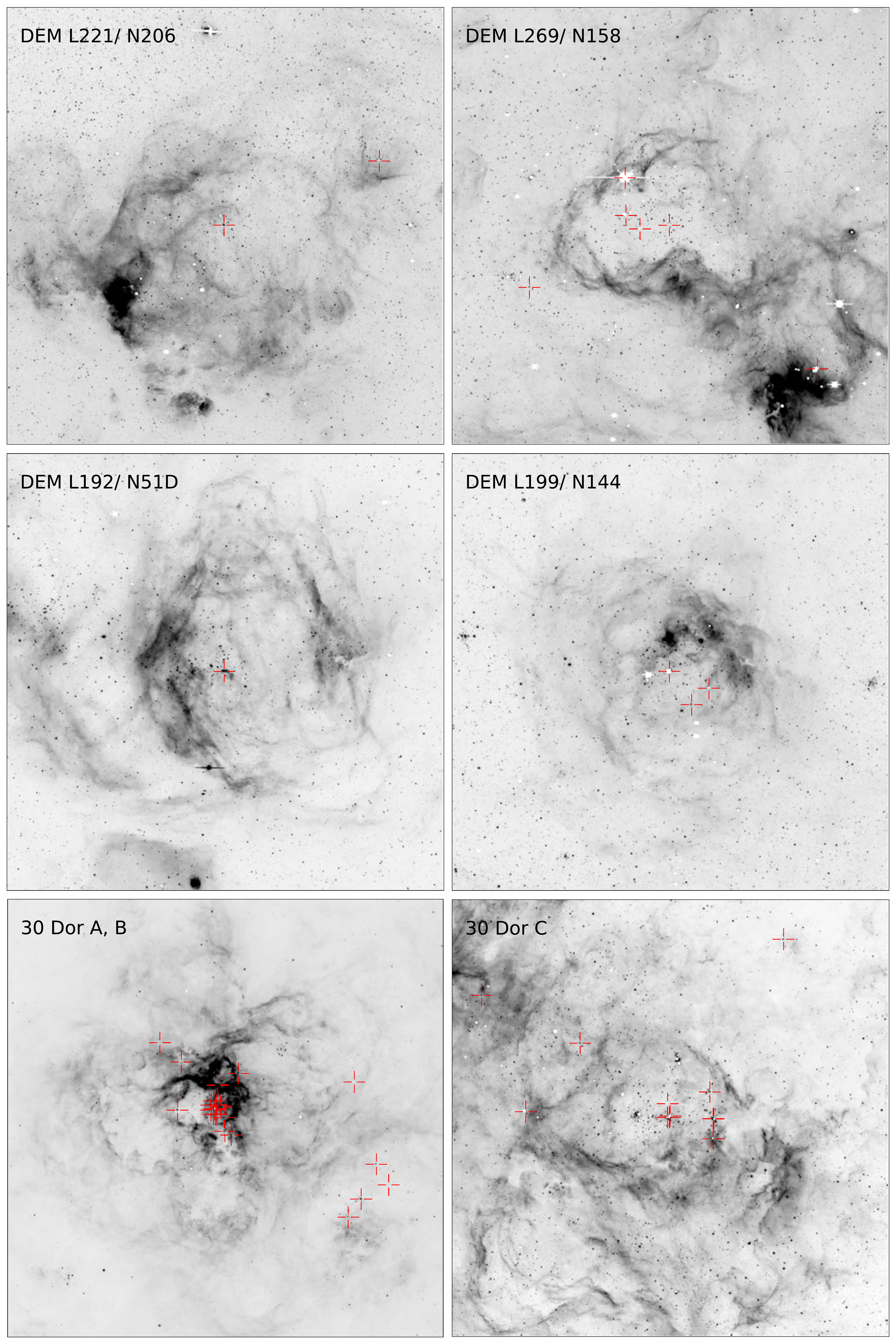}
\caption{4m MOSAIC H$\alpha$ images of example superbubbles around WR stars in the LMC. The field of view of each panel is 15'$\times$15'. The WR stars are marked by red crosses.}
\label{fig:superbubble}
\end{figure*}

\section{Discussion} 

Below we first examine the WR star population in the LMC, using their stellar and interstellar environments to assess their properties, then correlate their bubbles and superbubbles with their spectral types to further probe the nature of the WR spectral types.  Finally, we note the impact of WR nebulae on the supernova remnants that are formed after the supernova explosions. 

\subsection{WR Star Population in the LMC}

\subsubsection{30 Dor versus the Rest of the LMC}
The distribution of WR stars in the LMC is by no means uniform.  Most conspicuously, about 30\% of the WR stars in the LMC are concentrated in the 30 Dor giant \ion{H}{2} region.  As shown in Table 5, the largest difference in the WR star population inside 30 Dor and outside 30 Dor is in the number ratio of WN2-4 to WN5-6 stars, roughly 1:3 in 30 Dor and 8:1 in the rest of the LMC.  This reversal in the number ratio of WN2-4 to WN5-6 stars is caused by 30 Dor's young age and massive clusters, where the most massive bins of the initial mass function can be stochastically populated.  The evolution of massive stars has been reviewed by \citet{Langer2012}, whose Figure 10  shows that  the most massive stars evolve into late-type WN stars even during the core hydrogen-burning stage.  The large number of WN5-6 stars in 30 Dor coexist with the numerous very massive O2-3 stars \citep{Crowther2010,Crowther2016}, highly suggestive that these WN stars are very massive hydrogen-burning stars, which have been shown to be very luminous and still have hydrogen on their surface \citep{deKoter1997,Massey1998}.

\subsubsection{WR Stars and OB Associations}
The starburst in 30 Dor has produced massive star clusters or OB associations, namely, LH 100 (R136 cluster) in 30 Dor A \citep{Hunter1995,Massey1998}, LH 99 in 30 Dor B \citep{Chu1997}, and LH 90 in 30 Dor C \citep{Lortet1984,Testor1993}.  It is thus not surprising that 70\% of the WR stars in 30 Dor are in OB associations. Among different types of WR stars in 30 Dor, WN5-6 has a higher percentage of being in OB associations than WN2-4, WN7-L, and WC.  The R136 cluster is mostly responsible for this high percentage of WN5-6 stars in OB associations in 30 Dor, and these WN5-6 stars are most likely massive hydrogen-burning stars.

In contrast, only 45\% of WR stars outside 30 Dor are in OB associations (see Table 5).  Interestingly, about 87\% of the WC stars outside 30 Dor are in OB associations.  In 30 Dor, four out of seven WC stars are in OB associations, corresponding to 57\%; however, this is small number statistics.  Furthermore, two WC stars in 30 Dor are within 50 pc from OB associations and could be associated; thus, we do not think the percentage of WC stars in OB associations differs much between 30 Dor and outside 30 Dor.

\subsubsection{WR Stars and Surrounding Ionized ISM}

Massive stars inject energy into the ambient ISM via UV radiation and fast stellar winds during their lifetime and supernova explosions at the end.  The energy feedback is expected to produce an amorphous \ion{H}{2} region when the massive stars are very young.  Gradually, the fast stellar winds of massive stars sweep up the \ion{H}{2} region into a shell structure called a bubble (for an isolated massive star) or superbubble (for OB association).  Eventually, the shell structure dissipates into the ISM, leaving behind a low-density ISM.  These three stages correspond to the three morphological classes of \ion{H}{2} environments defined in Section 3.2, with class 1 being the youngest and class 3 being the most evolved stage.   

Table 7 shows that in 30 Dor, 15\% (7/46) of WR stars are superposed on dense ionized gas (class 1), over 72\% (33/46) of WR stars are associated with superbubbles or shell-like structures (class 2), and 13\% (6/46) of WR stars in the outskirts of 30 Dor are superposed on diffuse nebulosity with low surface brightness (class 3). 
Among the 33 WR stars in class 2 \hii regions, 29 are inside superbubbles, 3 are projected on the southern rim of the superbubble Shell 5, and 1 is inside a faint shell structure.  Twenty-eight of the 29 WR stars in superbubbles are in the two major OB associations (LH 90 and LH 100) that are both surrounded by superbubbles.  

Outside 30 Dor, 5\% (6/110) WR stars are associated with \ion{H}{2} class 1, 36\% (40/110) are in superbubbles, and 58\% (64/110) are in a very tenuous medium with density lower than 1 H-atom cm$^{-3}$.  It is interesting to note that WN2-4 and WN7-L have the highest percentages associated with \ion{H}{2} class 3, the most evolved state.  Based on the massive star evolution summarized in Figure 10 of \citet{Langer2012}, we suggest that these WN2-4 stars in a very tenuous medium have progenitors with 20--30 $M_\odot$ initial masses, and that these WN7-L stars are the massive helium-burning ones whose progenitors have $\ge 50 M_\odot$ initial masses. Note, however, that this conclusion can change if stellar rotation, close binary evolution, and magnetic fields are considered in models of massive star evolution \citep{Meynet2017}.

\subsection{WR Bubbles: Observation versus Expectation}

\subsubsection{Too Few WR Bubbles are Detected}

According to our current understanding of nebulae around WR stars, as detailed in Section 2.2, we may expect every WR star to be surrounded by a small circumstellar bubble and enclosed by a larger interstellar bubble. However, only six WR stars (WR 2, 4, 17, 52, 61, 154) are inside nested small and large shells, which could be candidates for circumstellar and interstellar bubbles. In fact, only 12\% of WR stars are surrounded by small bubbles and only a small fraction of these bubbles have abundance observations to confirm that they are in fact circumstellar bubbles.

The scarcity of small bubbles observed around WR stars needs explanation. It is possible that a close binary companion may perturb and redirect the mass outflow of the WR star's progenitor, prohibiting the formation of a circumstellar bubble. However, $\sim$24\% of LMC WR stars with small bubbles are binary, and among all LMC WR stars, $\sim$22\% are binary, not too different from that for WR stars with small bubbles, suggesting that binarity may not be a main factor in prohibiting the formation of circumstellar bubbles.

A more plausible cause for the nondetection of circumstellar bubbles may be in its evolution. It is conceivable that as a circumstellar bubble expands, its density decreases and its emission measure rapidly drops below the detection limit. (Emission measure $\equiv$ $n^2L$, where $n$ is the electron density and $L$ is the emitting path length). The maximum emission measure of a shell is along the path length tangent to the shell's inner rim. For a uniform shell with mass $M$, radius $R$, and fractional shell thickness $\Delta R / R$, the maximum emission measure will be $n^2L = 32^{-1/2} \pi^{-2} R^{-5} (\Delta R / R)^{-3/2} (M/m_H)^2$, where $m_H$ is the mass of a hydrogen atom. Using observer-friendly units, the emission measure can be expressed in units of cm$^{-6}$\,pc as $n^2L$ $\approx$ 30$R^{-5} (\Delta R / R)^{-3/2} M^2$, where $n$ is in units of cm$^{-3}$, $L$ in parsecs, $R$ in parsecs, and $M$ in $M_\odot$.

\begin{figure*}[tbh]
\centering
\includegraphics[width=\columnwidth]{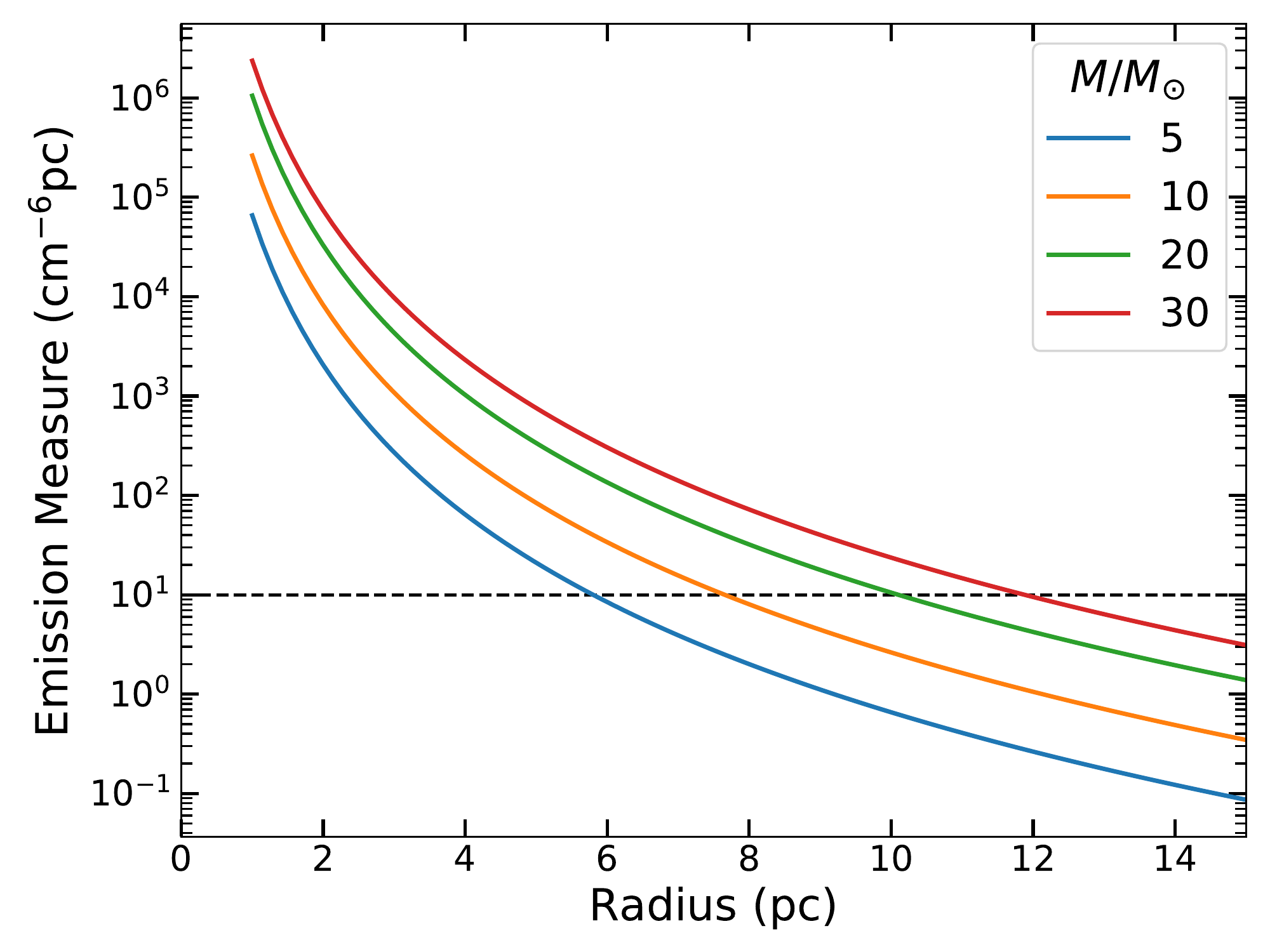}
\includegraphics[width=\columnwidth]{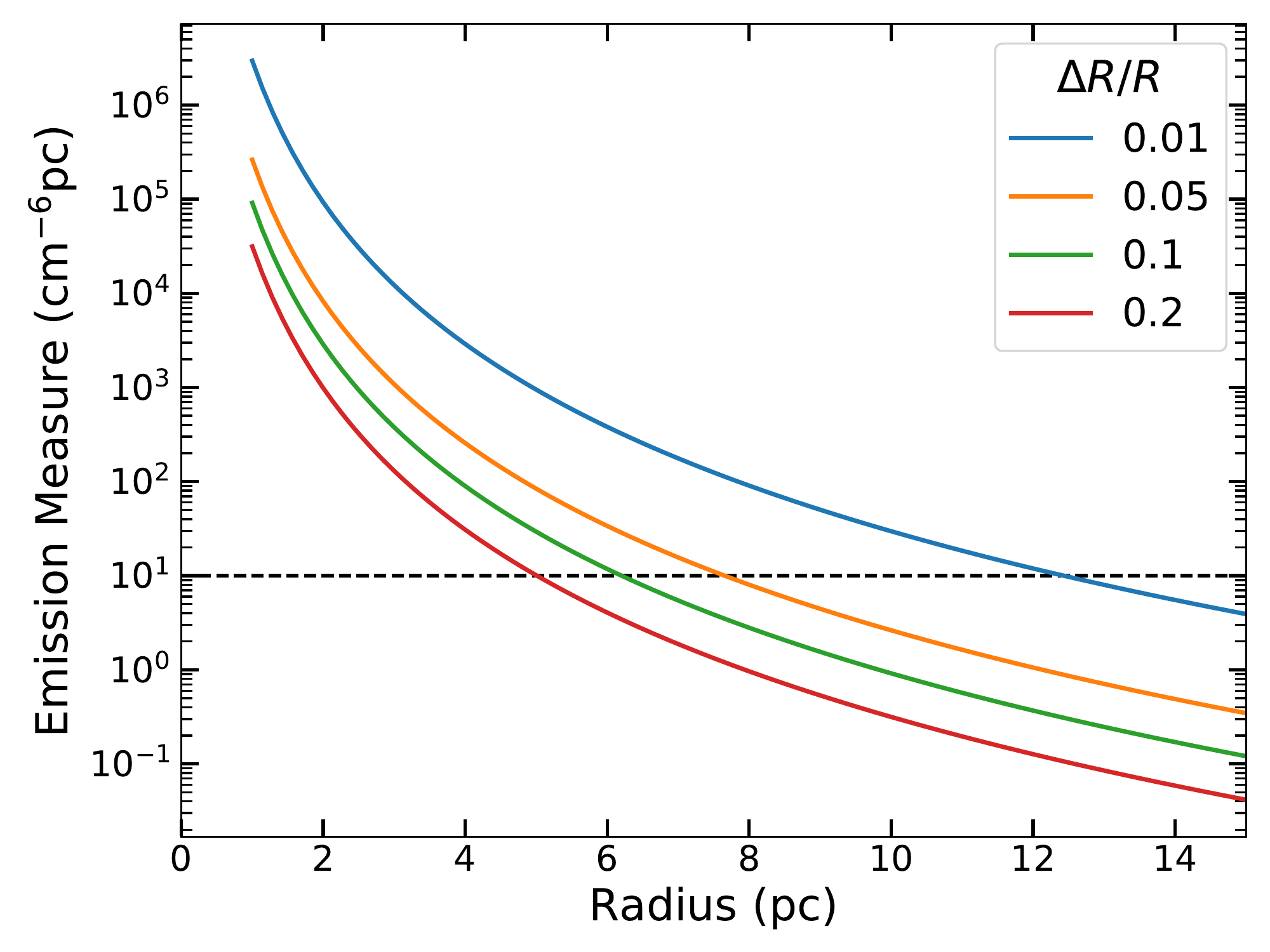}
\caption{Expected emission measures of circumstellar bubbles for a fixed shell thickness ratio $\Delta R/R=0.05$ and different bubble masses (left panel), and for $10\,M_{\odot}$ and different $\Delta R/R$ (right panel). The dashed line marks the emission measure 10 cm$^{-6}$ pc as the detection limit.}
\label{fig:EM}
\end{figure*}

It is evident that as the shell expands, the density drops and the maximum emission measure of the shell will decrease rapidly in proportion to $R^{-5}$. As shown in Figure~\ref{fig:EM}, for a 10\,$M_\odot$ shell with  $\Delta R / R$ = 0.05, the maximum emission measure drops below a detection limit of 10 cm$^{-6}$\,pc at a radius of $\sim$8 pc. Therefore, we do not expect to detect a circumstellar bubble around every WR star.

The invisible interstellar bubbles were a puzzle until \citet{Naze2003} showed that interstellar bubbles blown by MS O stars expand slowly, with expansion velocities of only 15--20 km~s$^{-1}$, producing very weak shocks and compression. Without a large density jump, the interstellar bubble cannot stand out against the background, and the bubble cannot be morphologically identified in direct images. Therefore, interstellar bubbles formed by MS stars are not expected to be seen around WR stars unless the WR star has gone through an RSG phase when the ambient ionized interstellar gas recombines and cools. The isothermal sound velocity of neutral \ion{H}{1} gas at 100 K is $\sim$1 km s$^{-1}$, and an interstellar bubble expanding at 15-20 \kms\ would generate strong shocks and compression, causing a large density jump. Such an interstellar bubble will become visible in \ha\ when the WR phase starts, and the bubble and ambient ISM become photoionized. Whether an interstellar bubble is visible around a WR star depends a lot on the evolutionary path and physical conditions of the ambient ISM.  

\subsubsection{Circumstellar versus Interstellar Bubbles}

The most definitive diagnostic of a circumstellar bubble is the N/O abundance ratio \citep[e.g.,][]{Esteban1991}.
Of the small WR bubbles identified, abundance observations have been made for only three--WR 2, 12, and 19--and only the bubble around WR 19 shows high N/O ratio confirming its circumstellar bubble nature \citep{Garnett1994,Stock2011}.  The small nebula near WR 47 (BAT99-38) was also observed by \citet{Stock2011}; however, this small nebula consists of bright-rimmed Bok globules with embedded star formation \citep{Chu2005} and not a WR nebula, as described in more detail in the Appendix.

In Section 5.2.1 we have shown that circumstellar bubbles expand and fade below the detection limit when the bubble size is greater than $\sim$20 pc.  The smallest WR bubbles are thus more likely to consist of CSM; however, WR 2 and WR 19 have bubbles of similarly small sizes but only the latter shows enhanced N/O ratio.  Future spectroscopic observations of the small WR bubbles are needed to determine unambiguously whether they consist of CSM.

\subsubsection{Correlation with Stellar Properties and Environments}

The statistics of small WR bubbles is summarized in Table 6.  About 67\% of the small bubbles are found around WN2-4 stars, and more strikingly, none of these WN2-4 stars are in OB associations.  If all massive stars are born in clusters or OB associations, the absence of OB associations around these WN2-4 stars with small bubbles suggests that the parent OB associations have dissolved after losing the more massive stars in supernova explosions and that these WN2-4 stars must have lower initial masses, $\sim$25 $M_\odot$. Among WN2-4 stars in superbubbles, only $\sim$50\% are in OB associations.  This superbubble statistic for WN2-4 stars might be caused by two possible evolutionary paths, with the ones in OB associations more massive than those not in OB associations.

In contrast, two WC4 stars are in small bubbles, and both are in OB associations.  This may look like small number statistics, but all 12 WC stars in superbubbles are in OB associations as well. The implication of ``WC stars in bubbles and superbubbles are all in OB associations'' is intriguing but not clear.

\subsection{Preexisting Environmental Conditions of Supernova Remnants}

Massive as they are, WR stars may explode as core-collapse supernovae near the end of their evolution. We have shown through this survey of bubbles around WR stars that circumstellar bubbles can be observed only at a young age, and as they expand, the density drops and the bubble fades to oblivion.  While the interstellar and circumstellar bubbles are not always detected, their shell densities are still above those of the diffuse ISM. Thus, when the central star eventually explodes as a supernova, the supernova ejecta quickly expands in the low-density bubble interior until it hits the circumstellar or interstellar bubble shell.  This is called cavity explosion. Supernova remnants formed through cavity explosion include N132D and N63A \citep{Hughes1987,Hughes1998}, and they are characterized by their large size and high X-ray luminosity.

\section{Summary and Conclusion} 

We have conducted a multiwavelength survey of ISM and CSM around WR stars in the LMC.  The latest LMC WR catalog by \citet{Neugent2018} contains 154 entries and should be fairly complete.  A number of survey data sets are available for us to probe the ionized and neutral medium around the WR stars: the MCELS images in the \ha, \oiii, and \sii lines; Blanco 4m Telescope MOSAIC images in the \ha\ line; \emph{Spitzer Space Telescope} 8 and 24 $\mu$m images; and ATCA + Parkes Telescope \hi 21 cm line data cube of the LMC.  

As a massive star evolves from the MS to WR phase, its fast MS wind turns into a slow RSG or LBV wind and then a fast WR wind.  The interactions between these winds and the ambient medium should form a circumstellar bubble inside an interstellar bubble \citep{Garcia1996II, Garcia1996I}.  We expect the interstellar and circumstellar bubbles to be small, $<$50 pc in size.  Our search finds small bubbles around $\sim$12\% of WR stars in the LMC, and around $\sim$15\% of WR stars in the LMC but outside 30 Dor.  Most of these small bubbles are around WN stars and most of them are not in OB associations.  Only the small bubble around WR 19 has a high N/O abundance ratio to confirm its circumstellar bubble nature.  The scarcity of small bubbles can be caused by several factors.  As a bubble expands, its surface brightness drops in proportion to (radius)$^{-5}$ and fades below the detection limit within a short time.  Furthermore, it has been observed that interstellar bubbles expand at 15--20 km s$^{-1}$, which cannot produce strong compression for the bubble shell to stand out against the background; thus, interstellar bubbles can be diagnosed kinematically but not morphologically \citep{Naze2003}.

WR stars and nearby massive stars can jointly blow a superbubble with their fast stellar winds and supernova explosions.  In 30 Dor, 63\% of its 46 WR stars are in superbubbles, and 28 of the 29 WR stars in superbubbles are also in OB associations. Outside 30 Dor, only 33\% of the 110 WR stars are in superbubbles; about two-thirds of these WR stars in superbubbles are also in OB associations. The contrasting statistics is caused by the prominent starburst in 30 Dor that produced the massive OB association LH 90 in 30 Dor C and more impressively the super star cluster R136 (LH 100) in the central superbubble of 30 Dor A.  While both inside and outside 30 Dor WN stars contribute to $\sim$80\% of the WR population, the number ratio of WN2-4 to WN5-6 stars is 1:3 in 30 Dor and 8:1 outside 30 Dor.  The dominant WN4-5 stars in 30 Dor are probably massive H-burning stars. 

We have also classified the \hii environments of WR stars into three categories: class 1 for bright amorphous \hii regions; class 2 for shell \hii regions, such as superbubbles; and class 3 for low-density diffuse medium.  These three classes correspond to a progression in the dynamical dispersal of ISM around massive stars.  Inside 30 Dor, the distribution of WR stars in classes 1-3 is 15\%, 72\%, and 13\%, respectively.  The stars in class 3 \hii environments are in the outskirts of 30 Dor.  Outside 30 Dor, the distribution of WR stars in classes 1--3 is 5\%, 36\%, and 58\%, respectively.  The contrast between 30 Dor and elsewhere reflects the age and mass differences in the WR star population.  For example, among the 65 WN2-4 stars outside 30 Dor, about 68\% are not in OB associations and 63\% are in class 3 \hii environments.  The WN2-4 stars not in OB associations and in class 3 \hii environments are most likely He-burning WR stars from progenitors with initial masses below $\sim$30 $M_\odot$.  

Finally, spectroscopic analyses of abundances are needed to determine whether the small WR bubbles contain ISM or CSM.  It is important to determine the physical properties and nature of small WR bubbles, as they will define how the future supernova remnants develop.
 
\acknowledgments 
C.S.H. acknowledges the hospitality of ASIAA for hosting her for the duration of the research. We acknowledge grant support of MOST 108-2112-M-001-045 and MOST 108-2811-M-001-58 from the Ministry of Science and Technology of Taiwan, Republic of China. 
 
\clearpage

\begin{longrotatetable}
\begin{deluxetable*}{cccccccccccccc}
\tablecaption{LMC WR Stars and Their Stellar and Interstellar Environments}
\tablehead{
            WR \# & $\alpha$(J2000) & $\delta$(J2000) & BAT99 & Brey & Spectral & OB assc. \tablenotemark{a}  & Nebula & Nebula & \hii & Bubble & Superbubble & \hii SGS & \hi Shell \tablenotemark{b} \\
             & & & Num. & Num. & Type & LH Num. & Henize N & DEM\,L & Morphology & (arcmin) & (arcmin) & LMC- & \\
             (1) & (2) & (3) & (4) & (5) & (6) & (7) & (8) & (9) & (10) & (11) & (12) & (13) & (14)} 
\startdata
            1   &  04 45 32.25  & -70 15 10.9  & 1   & 1   & WN3                      &...  & ... & ...  & 3c & ... & ...  & ... & ... \\
            2   &  04 49 36.25  & -69 20 54.8  & 2   & 2   & WN2                      &...  & 79  & 6    & 2a & 0.4 $\times$ 0.3 & 3.2 $\times$ 2.2 & ... & ...\\
            3   &  04 52 57.39  & -66 41 13.5  & 3   & 3   & WN3                      &... & ... & ...   & 3c & ... & ... & ... & ... \\
            4   &  04 53 03.76  & -69 23 51.8  &...  &...  & WN3+O6V                  & ((2)) & ... & 10 & 2a & 0.6 $\times$ 0.6 & 4.3 $\times$ 4 & 7 & GS\,9 \\
            5   &  04 53 30.08  & -69 17 49.4  & 4   & 3a  & WNL/Of?                  & ((5)) & ... & ... & 3b & ... & 18 $\times$ 12 & 7    & GS\,9 \\
            6   &  04 54 28.10  & -69 12 50.9  & 5   & 4   & WN2                      & 5   & 83  &  22  & 3a & ... & ... & 7    & GS\,15 \\
            7   &  04 55 07.60  & -69 12 31.7  & 5a  &...  & WN3                      & ((5)) & ... & 22 & 3a & ... & ... & 7    & GS\,15 \\
            8   &  04 55 31.35  & -67 30 02.7  & 7   &...  & WN3pec                   &...  & ... & ...  & 3b & ... & 11 $\times$ 7.6 & ... & ...\\
            9   &  04 56 02.88  & -69 27 21.5  & 8   & 8   & WC4                      & 8   & 94  & 36  & 2b & ... & 7.6 $\times$ 6.6 & 7    & GS\,18\\
            10  &  04 56 11.0   & -66 17 33.0  & 9   & 7   & WC4                      &...  & 11  &  34  & 2a & ... & (4 $\times$ 2) & ... & ... \\
            11  &  04 56 34.63  & -66 28 26.4  & 10  & 9   & WC4+OB                   & 9   & 11  &  34  & 2a & ... & 12 $\times$ 8 & ... & GS\,16 \\
            12  &  04 57 24.10  & -68 23 57.3  & 11  & 10  & WC4                      & 12  & 91  &  39  & 1/2a & 3 $\times$ 2 & ... & 6    & SGS\,2 \\
            13  &  04 56 48.79  & -69 36 40.7  &...  &...  & WN4/O4                   & ... & ... & ...  & 3c & ... & ... & 7    & GS\,18 \\
            14  &  04 57 27.44  & -67 39 02.9  & 12  &...  & O2If*/WN5                &...  & ... &  40  & 3a & 2.2 $\times$ 1.4 & ... & ... & ... \\
            15  &  04 57 41.04  & -66 32 42.6  & 13  &...  & WN10                     &...  & 11  &  34  & 3a & ... & ... & ... & ... \\
            16  &  04 58 56.36  & -68 48 04.7  & 14  & 11  & WN4+OB                   &...  & ... & ...  & 3c & ... & ... & 6    & SGS\,2 \\
            17  &  04 59 51.55  & -67 56 55.4  & 15  & 12  & WN3                      &...  & 16A &  45  & 2a & 2 $\times$ 2 & 10.2 $\times$ 8.4 & ... & ... \\
            18  &  05 02 59.24  & -69 14 02.3  & 15a &...  & WN3+abs                  &...  & ... & ...  & 3b & ... & 20 $\times$ 20 & ... & ... \\
            19  &  05 03 08.91  & -66 40 57.5  & 16  & 13  & WN7h                     &...  & ... &   5  & 3a & 0.7 $\times$ 0.3 & ... & ... & ... \\
            20  &  05 04 12.33  & -70 03 55.4  & 17  & 14  & WN4                      &...  & ... &  68  & 3a/3b & ... & 7.7 $\times$ 4.3 & 8    & ...\\
            21  &  05 04 32.64  & -68 00 59.4  &...  &...  & WN3/O3                   &...  & 23  &  66  & 2a    & 3 $\times$ 2.8 & ...    & ... & SGS\,5\\
            22  &  05 05 08.43  & -70 22 45.1  & 18  & 15  & WN3h                     & ... & ... &  80  & 3c & ... & ... & 8    & ...\\
            23  &  05 07 13.33  & -70 33 33.9  &...  &...  & WN3/O3                   &((26)) & ... &  80  & 3a & ... & ... & 8    & ...\\
            24  &  05 09 40.42  & -68 53 24.8  & 19  & 16  & WN3+OB                   & 31  &105A &  86  & 2a & ... &  4.4 $\times$ 3 & ... & SGS\,5\\
            25  &  05 09 53.77  & -68 52 52.5  & 20  & 16a & WC4                      & 31  &105A &  86  & 1/2a & 2.8 $\times$ 1.4 & ... & ... & SGS\,5\\
            26  &  05 13 43.77  & -67 22 29.4  & 21  & 17  & WN4+OB                   & 37  & 30  & 105  & (2b) & ... & (10.2 $\times$ 7.4) & ... & GS\,44\\
            27  &  05 13 54.27  & -69 31 46.7  & 22  & 18  & WN9                      & 39  & ... & 110  & 2b & ... & 7 $\times$ 4.6 & ... & GS\,45\\
            28  &  05 13 56.01  & -67 24 36.6  & 23  & ... & WN3                      &(37)/(38) & 30  & 105  & 3a &  ... & ... & ... & GS\,44\\
            29  &  05 14 12.69  & -69 19 26.2  & 24  & 19  & WN3                      &(35) & 113 & 108  & 2a & 2.8 $\times$ 2.6 & ... & ... & GS\,46\\
            30  &  05 14 17.57  & -67 20 35.1  &...  & ... & O3.5If*/WN5              &(36) & 30  & 105  & 2b & ... & 10.2 $\times$ 7.4 & ... & GS\,44\\
            31  &  05 14 57.27  & -71 36 18.3  & 25  & 19a & WN4ha                    &...  & ... & 119  & 3a & ... & ... & ... & GS\,47\\
            32  &  05 16 38.84  & -69 16 40.9  & 26  & 20  & WN4                      &...  & 119 & ...  & 3a/3c & ... & ... & ... & GS\,51\\
            33  &  05 18 10.33  & -69 13 02.5  &...  &...  & WO4                      & 41  & 119 & 132a & 3a & ... & ... & ... & GS\,54\\
            34  &  05 18 10.88  & -69 13 11.4  &...  &...  & WO4                      & 41  & 119 & 132a & 3a & ... & ... & ... & GS\,54\\
            35  &  05 18 19.21  & -69 11 40.7  & 27  & 21  & BI+WN4                   & 41  & 119 & 132b & 3a/3b & ... & ... & ... & GS\,54\\
            36  &  05 19 16.34  & -69 39 20.0  & 28  & 22  & WC6+O5-6                 & 42  & 120 & 134  & 3a & ... & ... & ... & ...\\
            37  &  05 20 44.73  & -65 28 20.5  & 29  & 23  & WN3+OB                   & 43  & ... & 137  & 2b & ... & 13.8 $\times$ 12.8 & ... & SGS\,6 \\
            38  &  05 21 22.82  & -65 52 48.8  &...  & ... & WN3/O3                   & 45  & ... & 154  & 3a/3b & ... & ... & 5    & SGS\,7\\
            39  &  05 21 57.70  & -65 49 00.3  & 30  & 24  & WN6h                     & 45  & ... & 154  & 3b & ... & ...  & 5    & SGS\,7\\
            40  &  05 22 04.41  & -67 59 06.8  & 31  & 25  & WN3                      & 47  & 44C & 152  & 1 & ... & ... & ... & ...\\
            41  &  05 22 22.53  & -71 35 58.1  & 32  & 26  & WN6h                     &...  & 198 & 165  & 2b & ... & 8.1 $\times$ 5.5 & ... & GS\,61\\
            42  &  05 22 59.78  & -68 01 46.6  & 33  &...  & Ofpe/WN9                 & 49  & 44  & 160  & 1 & ... & ... & ... & ... \\
            43  &  05 23 10.06  & -71 20 50.9  & 34  & 28  & WC4+abs                  & 50  & 200 & 164  & 2b & ... & 15.4 $\times$ 12.7 & 9    & GS\,61\\
            44  &  05 23 18.01  & -65 56 57.0  & 35  & 27  & WN3                      &...  & ... & 154  & 3c & ... & ... & 5    & SGS\,7\\
            45  &  05 24 24.19  & -68 31 35.6  & 36  & 29  & WN3/WCE+OB               &...  & 138 & 174  & 1 & ... & ... & 3    & ... \\
            46  &  05 24 54.34  & -66 14 11.1  & 37  & 30  & WN3                      &((52)) & ... & 175 & 2b &  ... & 7.6 $\times$ 3.8 & 5    & SGS\,7\\
            47  &  05 26 03.96  & -67 29 57.1  & 38  & 31  & WC4+abs                  & 54  & 51D & 192  & 2b & ... & 12 $\times$ 9.5 & 4 & SGS\,11\\
            48  &  05 24 56.87  & -66 26 44.4  &...	 &...  & WN3/O3                   &...  & 48E & 175a & 2b & (3.8 $\times$ 2.9) & ... & 4, 5 & SGS\,7, 11\\
            49  &  05 26 30.26  & -68 50 27.5  & 39  & 32  & WC4+O6III/V              & 58  & 144 & 199  & 2a & ... & 10 $\times$ 10 & 3    & SGS\,12\\
            50  &  05 26 36.86  & -68 51 01.3  & 40  & 33  & WN4                      & 58  & 144 & 199  & 2a & ... & 10 $\times$ 10 & 3    & SGS\,12\\
            51  &  05 26 42.58  & -69 06 57.4  & 41  & 35  & WN4                      &...  & 135 & 198  & 3a & ... & ... & 3    & SGS\,12\\
            52  &  05 26 45.32  & -68 49 52.8  & 42  & 34  & B3I+WN5                  & 58  & 144 & 199  & 2a & 0.8 $\times$ 0.8 & 10 $\times$ 10 & 3    & SGS\,12\\
            53  &  05 27 37.68  & -70 36 05.4  & 43  & 37  & WN3+OB                   & 62  & 204 & 208  & 2a/2b & ... & 14.4 $\times$ 13.5 & 9    & ...\\
            54  &  05 27 42.69  & -69 10 00.4  & 44  & 36  & WN7h                     &...  & 135 & 210  & 3b & 3.0 $\times$ 2.4 & ... & 3    & SGS\,12\\
            55  &  05 27 52.66  & -68 59 08.5  & 45  &...  & WN10$\rightarrow$LBV     & 61  & 135 & 210  & 3a & ... & ... & 3    & SGS\,12\\
            56  &  05 28 17.90  & -69 02 35.9  & 46  & 38  & WN4                      &((61)) & 135 & 210 & 3a &  ... & ... & 3    & SGS\,12\\    
            57  &  05 28 27.12  & -69 06 36.2  &...  &...  & WN3/O3                   &...  & 135 & 210  & 3a & ... & ... & 3    & SGS\,12\\
            58  &  05 29 12.37  & -68 45 36.1  & 47  & 39  & WN3                      & 64  & 135 & ...  & 3c & ... & ... & 3    & SGS\,12\\
            59  &  05 29 18.19  & -69 19 43.2  &...  &...  & WN3/O3                   &...  & 135 & ...  & 3c & ... & ... & 3    & SGS\,12\\
            60  &  05 29 31.64  & -68 54 28.8  & 48  & 40  & WN3                      &...  & 135 & ...  & 3c & ... & ... & 3    & SGS\,12\\
            61  &  05 29 33.21  & -70 59 34.9  & 49  & 40a & WN3+O7.5                 &((66)) & 206 & 221  & 2a & 1.4 $\times$ 1.4 & 14 $\times$ 11 & 9    & GS\,70\\
            62  &  05 29 53.64  & -69 01 04.8  & 50  & 41  & WN5h                     &...  & 135 & ...  & 3c & ... & ... & 3    & SGS\,12\\
            63  &  05 30 02.46  & -68 45 18.4  & 51  & 42  & WN 3                     &((64)) & 135 & 218 & 3c &  ... & ... & 3    & SGS\,12\\
            64  &  05 30 12.16  & -67 26 08.3  & 52  & 43  & WC4                      &...  & ... & ...  & 3c & ... & ... & 4    & SGS\,11\\
            65  &  05 30 38.70  & -71 01 47.8  & 53  & 44  & WC4+abs                  & 69  & 206 & 221  & 2a & ... & 14 $\times$ 11 & 9    & GS\,70\\
            66  &  05 31 18.05  & -69 08 45.5  & 54  &...  & WN9                      & ... & 135 & 232  & 3a & ... & ... & 3    & SGS\,12\\
            67  &  05 31 25.52  & -69 05 38.6  & 55  &...  & WN11                     &...  & 135 & 232  & 3a & ... & ... & 3    & SGS\,12\\
            68  &  05 31 32.87  & -67 40 46.6  & 56  & 46  & WN3                      & ((76)) & 57 & 229 & 2a & ... & 11.5 $\times$ 6.5 & 4 & ...\\
            69  &  05 31 34.36  & -67 16 29.3  & 57  & 45  & WN3                      &((70)) & ... & ... & 3a & ... & ... & 4    & SGS\,11\\
            70  &  05 32 07.49  & -68 26 31.6  & 58  & 47  & WN7h                     &...  & 148 & 227  & 3a & ... & ... & 3 & SGS\,12\\
            71  &  05 33 10.57  & -67 42 43.1  & 59  & 48  & WN3+OB                   &((76)) & 57C & 231 & 2b &  1.5 $\times$ 1.2 & ... & ... & ...\\
            72  &  05 33 10.87  & -69 29 01.0  & 60  & 49  & WN3+OB                   &...  & 135 & 224  & 3a &  ... & ... & ... & ...\\
            73  &  05 34 19.24  & -69 45 10.3  & 61  & 50  & WC4                      & 81  & 154 & 246  & 2a & ... & 21 $\times$ 15 & ... & GS\,73\\
            74  &  05 34 36.08  & -69 45 36.5  &...  &...  & B0I+WN                   & 81  & 154 & 246  & 2a & ... & 21 $\times$ 15 & ... & GS\,73\\
            75  &  05 34 37.47  & -66 14 38.0  & 62  & 51  & WN3                      &...  & 62A & 239  & 3a & 0.7 $\times$ 0.2 & ... & 4    & SGS\,11\\
            76  &  05 34 52.03  & -67 21 29.0  & 63  & 52  & WN4h                     &...  & ... & 240  & 3a & 1.5 $\times$ 1.1 & ... & 4    & SGS\,11\\
            77  &  05 34 59.38  & -69 44 06.3  & 64  & 53  & WN3+O                    & 81  & 154 & 246  & 2a & ... & 21 $\times$ 15 & ... & GS\,73\\
            78  &  05 35 00.90  & -69 21 20.2  &...  &...  & WN3/O3                   &...  & 157 & 263  & 3a & ... & ... & ... & ...\\
            79  &  05 35 15.18  & -69 05 43.1  & 65  & 55  & WN3                      &((89)) & 157 & 263 & 3a & 0.5 $\times$ 0.3 & ... & ... & ...\\
            80  &  05 35 28.52  & -69 40 08.9  &...  &...  & WN3+O8-9III              & 87  & 154 & 246  & 2a & ... & 21 $\times$ 15 & ... & GS\,73\\
            81  &  05 35 29.80  & -67 06 49.4  & 66  & 54  & WN3(h)                   &...  & ... & ...  & 3a &  ... & ... & 4   & SGS\,11\\
            82  &  05 35 41.96  & -69 11 52.9  & 69  &...  & WC4                      & 90  & 157C& 263  & 2b & ... & 7.5 $\times$ 5.2 & ... & GS\,75\\
            83  &  05 35 42.19  & -69 12 34.5  & 67  & 56  & WN5h                     & 90  & 157C& 263  & 2b & ... & 7.5 $\times$ 5.2 & ... & GS\,75\\
            84  &  05 35 42.20  & -69 11 53.6  & 68  & 58  & O3.5If*/WN7              & 90  & 157C& 263  & 2b & ... & 7.5 $\times$ 5.2 & ... & GS\,75\\
            85  &  05 35 43.49  & -69 10 58.0  & 70  & 62  & WC4                      & 90  & 157C& 263  & 2b & ... & 7.5 $\times$ 5.2 & ... & GS\,75\\
            86  &  05 35 44.28  & -68 59 36.8  & 71  & 60  & WN3+abs                  & 89  & 135 & ...  & 3a & ... & ... & 3 & ...\\
            87  &  05 35 45.03  & -68 58 44.4  & 72  & 61  & WN4+abs                  & 89  & 135 & ...  & 3a & ... & ... & 3 & ...\\
            88  &  05 35 50.65  & -68 53 39.2  & 73  & 63  & WN5                      & 85  & 135 & ...  & 3a & ... & ... & 3 & ...\\
            89  &  05 35 52.43  & -68 55 08.7  & 74  & 63a & WN3+abs                  & 89  & 135 & ...  & 3a & ... & ... & 3 & ...\\
            90  &  05 35 54.03  & -67 02 48.9  & 75  & 59  & WN4                      &...  & ... & ...  & 3c & ... & ... & 4   & SGS\,11\\
            91  &  05 35 54.37  & -68 59 07.9  & 76  &...  & WN9                      & 89  & 135 & ...  & 3a & ... & ... & 3 & ...\\
            92  &  05 35 58.87  & -69 11 47.8  & 77  & 65  & WN7                      & 90  & 157C& 263  & 2b & ... & 7.5 $\times$ 5.2 & ... & GS\,75\\
            93  &  05 35 59.16  & -69 11 50.7  & 78  & 65b & WN4                      & 90  & 157C& 263  & 2b & ... & 7.5 $\times$ 5.2 & ... & GS\,75\\
            94  &  05 35 59.82  & -69 11 22.3  & 79  & 57  & WN7                      & 90  & 157C& 263  & 2b & ... & 7.5 $\times$ 5.2 & ... & GS\,75\\
            95  &  05 35 59.89  & -69 11 50.6  & 80  &...  & WN5                      & 90  & 157C& 263  & 2b & ... & 7.5 $\times$ 5.2 & ... & GS\,75\\
            96  &  05 36 12.13  & -67 34 57.8  & 81  & 65a & WN5h                     & 88  & 59B & 241  & 2a & ... & ... & ... & ...\\
            97  &  05 36 33.58  & -69 09 17.3  & 82  & 66  & WN3                      &(90) & 157 & 263  & 2b & 1.5 $\times$ 0.7 & (6.6 $\times$ 2.7) & ... & ...\\
            98  &  05 36 43.71  & -69 29 47.5  & 83  &...  & Ofpe/WN9$\rightarrow$LBV & 94  & 135 & 248  & 3c & ... & ... & ... & ...\\
            99  &  05 36 51.38  & -69 25 56.7  & 84  & 68  & WC4(+OB)                 & 96  & 135 & 261  & 3b & ... & 6.9 $\times$ 5.2 & ... & ...\\
            100 &  05 36 54.66  & -69 11 38.3  & 85  & 67  & WC4(+OB)                 &((90)) & 157 & 263 & 3a & ... & ... & ... & ...\\
            101 &  05 37 11.48  & -69 07 38.2  & 86  & 69  & WN3                      &(100)& 157A & 263  & 3a & ... & ... & ... & ...\\
            102 &  05 37 29.24  & -69 20 47.5  & 87  & 70  & WC4                      & 97  & 135 & 261  & 3a & ... & ... & ... & ...\\
            103 &  05 37 35.72  & -69 08 40.3  & 88  & 70a & WN3/WCE                  &(99) & 157A & 263  & 3a & ... & ... & ... & ...\\
            104 &  05 37 40.50  & -69 07 57.7  & 89  & 71  & WN6                      &(99) & 157A & 263  & 3a & ... & ... & ... & ...\\
            105 &  05 37 44.64  & -69 14 25.7  & 90  & 74  & WC4                      &(99) & 157 & 263  & 3a & ... & ... & ... & ...\\
            106 &  05 37 46.35  & -69 09 09.6  & 91  & 73  & WN6h                     & 99  & 157B & 263  & 3a & ... & ... & ... & ...\\
            107 &  05 37 47.62  & -69 21 13.6  &...  &...  & WN3+O7V                  & 97  & 135 & 261  & 3b & ...  & (4.3 $\times$ 1.6) & ... & ...\\
            108 &  05 37 49.04  & -69 05 08.3  & 92  & 72  & B1I+WN3                  & ((99,100)) & 157A & 263  & 2a & ... & 8 $\times$ 7 & ... & ...\\
            109 &  05 37 51.34  & -69 09 46.7  & 93  &...  & O3If*                    & 99  & 157B & 263  & 1 & ... & ... & ... & ...\\
            110 &  05 38 24.21  & -69 29 13.4  &...  &...  & WN11                     & 101 & 158 & 269  & 3a & ... & ... & ... & ...\\
            111 &  05 38 27.71  & -69 29 58.5  & 94  & 85  & WN3/4pec                 & 101 & 158 & 269  & 3a & ... & ... & ... & ...\\
            112 &  05 38 33.62  & -69 04 50.5  & 95  & 80  & WN7                      & (100) & 157A& 263 & 1 & ... & ... & ... & GS\,78\\
            113 &  05 38 36.42  & -69 06 57.4  & 96  & 81  & WN7                      & (100) & 157A& 263 & 1 & ... & ... & ... & GS\,78\\
            114 &  05 38 38.84  & -69 06 49.5  & 97  &...  & O3.5If*/WN7              & (100) & 157A& 263 & 1 & ... & ... & ... & GS\,78\\
            115 &  05 38 39.15  & -69 06 21.2  & 98  & 79  & WN6                      & 100 & 157A& 263  & 2a & ... & 2.5 $\times$ 1.4 & ... & GS\,78\\
            116 &  05 38 40.23  & -69 05 59.8  & 99  &...  & O2.5If*/WN6              & 100 & 157A& 263  & 2a & ... & 2.5 $\times$ 1.4 & ... & GS\,78\\
            117 &  05 38 40.55  & -69 05 57.1  & 100 &...  & WN6h                     & 100 & 157A& 263  & 2a & ... & 2.5 $\times$ 1.4 & ... & GS\,78\\
            118 &  05 38 41.60  & -69 05 14.0  & 101 &...  & WC4                      & 100 & 157A& 263  & 2b & (1.1 $\times$ 0.9)  & ... & ... & GS\,78\\
            119 &  05 38 41.60  & -69 05 14.0  & 102 &...  & WN6                      & 100 & 157A& 263  & 2b & (1.1 $\times$ 0.9) & ... & ... & GS\,78\\
            120 &  05 38 41.62  & -69 05 15.1  & 103 &...  & WN5(h)+O                 & 100 & 157A& 263  & 2b & (1.1 $\times$ 0.9) & ... & ... & GS\,78\\
            121 &  05 38 41.88  & -69 06 14.4  & 104 &...  & O2If*/WN5                & 100 & 157A& 263  & 2a & ... & 2.5 $\times$ 1.4 & ... & GS\,78\\
            122 &  05 38 42.12  & -69 05 55.2  & 105 &...  & O2If*                    & 100 & 157A& 263  & 2a & ... & 2.5 $\times$ 1.4 & ... & GS\,78\\
            123 &  05 38 42.33  & -69 06 03.3  & 106 &...  & WN4.5h                   & 100 & 157A& 263  & 2a & ... & 2.5 $\times$ 1.4 & ... & GS\,78\\
            124 &  05 38 42.39  & -69 06 02.9  & 108 &...  & WN5h                     & 100 & 157A& 263  & 2a & ... & 2.5 $\times$ 1.4 & ... & GS\,78\\
            125 &  05 38 42.41  & -69 06 15.0  &...  &...  & O2If*/WN5                & 100 & 157A& 263  & 2a & ... & 2.5 $\times$ 1.4 & ... & GS\,78\\
            126 &  05 38 42.41  & -69 06 02.9  & 109 &...  & WN5h                     & 100 & 157A& 263  & 2a & ... & 2.5 $\times$ 1.4 & ... & GS\,78\\
            127 &  05 38 42.43  & -69 06 02.7  & 110 &...  & O2If*/O3If*/WN6          & 100 & 157A& 263  & 2a & ... & 2.5 $\times$ 1.4 & ... & GS\,78\\
            128 &  05 38 42.91  & -69 06 04.8  & 112 &...  & WN4.5h                   & 100 & 157A& 263  & 2a & ... & 2.5 $\times$ 1.4 & ... & GS\,78\\
            129 &  05 38 43.10  & -69 05 46.8  & 113 &...  & O2If*/WN5                & 100 & 157A& 263  & 2a & ... & 2.5 $\times$ 1.4 & ... & GS\,78\\
            130 &  05 38 43.21  & -69 06 14.4  & 114 &...  & O2If*/WN5                & 100 & 157A& 263  & 2a & ... & 2.5 $\times$ 1.4 & ... & GS\,78\\
            131 &  05 38 44.06  & -69 05 55.6  & 115 &...  & WC5                      & 100 & 157A& 263  & 2a & ... & 2.5 $\times$ 1.4 & ... & GS\,78\\
            132 &  05 38 44.26  & -69 06 05.9  & 116 &...  & WN4.5h                   & 100 & 157A& 263  & 2a & ... & 2.5 $\times$ 1.4 & ... & GS\,78\\
            133 &  05 38 47.52  & -69 00 25.3  & 117 & 88  & WN4.5                    & 100 & 157A& 263  & 2a & ... & 8 $\times$ 4 & ... & ... \\
            134 &  05 38 53.38  & -69 02 00.9  & 118 &...  & WN5/6+WN6/7              & 100 & 157A& 263  & 2a & ... & 8 $\times$ 4 & ... & ... \\
            135 &  05 38 55.53  & -69 04 26.7  &...  &...  & WN5h                     & (100) & 157A& 263 & 1 & ... & ... & ... & GS\,78\\
            136 &  05 38 57.07  & -69 06 05.7  & 119 & 90  & WN6+O3.5If*/WN7          & 100 & 157A& 263  & 2a & ... & 2.5 $\times$ 1.4 & ... & GS\,78\\
            137 &  05 38 58.09  & -69 29 19.5  & 120 & 91  & WN9                      & 101 & 158 & 269  & 1 & ... & ... & ... & ...\\
            138 &  05 39 03.78  & -69 03 46.5  & 121 &...  & WC4                      & (100) & 157A& 263 & 1 & ... & ... & ... & ...\\
            139 &  05 39 11.33  & -69 02 01.6  & 122 & 92  & WN5h                     & ((100)) & 157A& 263  & 1 & ... & ... & ... & ... \\
            140 &  05 39 34.29  & -68 44 09.2  & 123 & 93  & WO3                      & ... & ... & 268  & 3a & ... & ... & ... & ...\\
            141 &  05 39 36.18  & -69 39 11.2  & 124 & 93a & WN3                      & 103 & 160 & 284  & 1 & ... & ... & 2    & SGS\,19\\
            142 &  05 39 56.11  & -69 24 24.3  & 125 & 94  & WC4+abs                  & 104 & 158 & 269  & 2a & ... & 7.5 $\times$ 6.3 & 2    & SGS\,19\\
            143 &  05 40 03.57  & -69 37 53.1  &...  &...  & WN3/O3                   & 103 & 160 & 284  & 2a & ... & 12 $\times$ 8 & 2    & SGS\,19\\
            144 &  05 40 07.55  & -69 24 31.9  & 126 & 95  & WN3+O7                   & 104 & 158 & 269  & 2a & ... & 7.5 $\times$ 6.3 & 2    & SGS\,19\\
            145 &  05 40 13.06  & -69 24 04.2  & 127 &...  & WC4+O                    & 104 & 158 & 269  & 2a & ... & 7.5 $\times$ 6.3 & 2    & SGS\,19 \\
            146 &  05 40 13.33  & -69 22 46.5  &...  &...  & B[e]+WN?                 & 104 & 158 & 269  & 2a & ... & 7.5 $\times$ 6.3 & 2    & SGS\,19\\
            147 &  05 40 50.80  & -69 26 31.8  & 128 & 96  & WN3                      &(104)& 158 & 269  & 3a & ... & ... & 2    & SGS\,19 \\
            148 &  05 41 17.50  & -69 06 56.2  &...  &...  & WN3/O3                   & ... & 135 & 310  & 3a & (2.8 $\times$ 1.8) & (6.3 $\times$ 4.5) & 2  & SGS\,19\\
            149 &  05 41 48.57  & -70 35 30.8  & 129 & 97  & WN3+O5V                  & ... & ... & 294  & 3a & ... & ... & ... & ...\\
            150 &  05 44 31.03  & -69 20 15.5  & 130 &...  & WN11                     & ... & 135 & 310  & 3a & ... & ... & 2    & SGS\,19 \\
            151 &  05 44 53.72  & -67 10 36.2  & 131 & 98  & WN4                      & 116 & 74  & 309  & 2a & ... & 17 $\times$ 15 & ... & GS\,94\\
            152 &  05 45 24.16  & -67 05 56.8  & 132 & 99  & WN4                      &((116))& ... & 309  & 3c & ... & ... & ... & ...\\
            153 &  05 45 51.93  & -67 14 25.9  & 133 &...  & WN11                     &((116))& ... & 308  & 2a & ... & 17 $\times$ 15 & ... & GS\,94 \\
            154 &  05 46 46.35  & -67 09 58.3  & 134 & 100 & WN3                      & ... & 74  & 315  & 2a & 1.8 $\times$ 1.3 & 7.5 $\times$ 5.2 & ... & GS\,94 
\enddata
\tablenotetext{a}{Single parentheses around OB association indicate that the star is within 50 pc of the center of the OB association, double parentheses indicate that the star is within 100 pc, and no parentheses indicate that the star is within the association. Ellipses indicate no association.}
\tablenotetext{b}{GS: giant shell, SGS: supergiant shell.}
\end{deluxetable*}
\end{longrotatetable}

\begin{deluxetable*}{cccccc}
    \tablecaption{WR Stars with Small Bubbles}
    \tablehead{WR \#  & Spectral Type & \multicolumn{2}{c}{Dimensions} & Nebula & OB \\ 
                 &  & (arcmin) & (pc) & DEM L & LH \\ 
                (1) & (2) & (3) & (4) & (5) & (6)}
        \startdata
            2 & WN2 & 0.4 $\times$ 0.3 & 6 $\times$ 4.5        &  6  & ... \\ 
            4 & WN3+O6V & 0.6 $\times$ 0.6 & 9 $\times$ 9      &  10 & ((2)) \\ 
            12 & WC4 & 3 $\times$ 2 & 45 $\times$ 30           &  39 & 12 \\ 
            14 & O2If*/WN5 & 2.2 $\times$ 1.4 & 33 $\times$ 21 & 40  &  ... \\ 
            17 & WN3 & 2 $\times$ 2 & 30 $\times$ 30           & 45  & ... \\ 
            19 & WN7h & 0.7 $\times$ 0.3 & 10.5 $\times$ 4.5   &  5  & ... \\ 
            21 & WN3/O3 & 3 $\times$ 2.8 & 45 $\times$ 42      & 66  & ... \\ 
            25 & WC4 & 2.8 $\times$ 1.4 & 42 $\times$ 21       & 86  & 31 \\ 
            29 & WN3 & 2.8 $\times$ 2.6 & 42 $\times$ 39       & 108 & (35) \\ 
            52 & B3I+WN5 & 0.8 $\times$ 0.8 & 12 $\times$ 12   & 199 & 58 \\ 
            54 & WN7h & 3.0 $\times$ 2.4 & 45 $\times$ 36      & 210 & ... \\ 
            61 & WN3+O7.5 & 1.4 $\times$ 1.4 & 21 $\times$ 21  & 221 & ((66)) \\ 
            71 & WN3+OB & 1.5 $\times$ 1.2 & 22.5 $\times$ 18  & 231 & ((76)) \\ 
            75 & WN3 & 0.7 $\times$ 0.2 & 10.5 $\times$ 3      & 239 & ... \\ 
            76 & WN4h & 1.5 $\times$ 1.1 & 22.5 $\times$ 16.5  & 240 & ... \\ 
            79 & WN3 & 0.5 $\times$ 0.3 & 7.5 $\times$ 4.5     & 263 & ((89)) \\ 
            154 & WN3 & 1.8 $\times$ 1.3 & 27 $\times$19.5     & 315 & ...\\ 
            \hline
            97 $\dagger$ & WN3 & 1.5 $\times$ 0.7 & 22.5 $\times$ 10.5  & 263 & (90)  
        \enddata
        \tablecomments{$\dagger$ In 30 Dor.}
\end{deluxetable*}

\begin{deluxetable*}{cccccc}
    \tablecaption{WR Stars in Superbubbles (Excluding 30 Dor)}
    \tablehead{WR \#  & Spectral Type & \multicolumn{2}{c}{Dimensions} & Nebula & OB \\ 
                 &  & (arcmin) & (pc) & DEM L & LH \\
                (1) & (2) & (3) & (4) & (5) & (6)}
        \startdata
            2 & WN2	& 3.2 $\times$ 2.2 & 48 $\times$ 33      &   6  & ...  \\ 
            4 & WN3+O6V & 4.3 $\times$ 4 & 64.5 $\times$ 60  &  10  & ((2)) \\ 
            5 & WNL/Of? & 18 $\times$ 12 & 270 $\times$ 180  & ...  & ((5))  \\ 
            8 & WN3pec	& 11 $\times$ 7.6 & 165 $\times$ 114 & ...  & ... \\ 
            9 & WN3 & 7.6 $\times$ 6.6 & 114 $\times$ 99     &  36  &  8 \\ 
            11 & WC4+OB & 12 $\times$ 8 & 180 $\times$ 120   &  34  &  9 \\ 
            17 & WN3 & 10.2 $\times$ 8.4 & 153 $\times$ 126  &  45  & ... \\ 
            18 & WN3+abs & 20 $\times$ 20 & 300 $\times$ 300  & ... & ... \\ 
            20 & WN4 & 7.7 $\times$ 4.3 & 115.5 $\times$ 64.5 &  68 & ... \\ 
            24 & WN3+OB & 4.4 $\times$ 3 & 66 $\times$ 45     &  86 & 31 \\ 
            27 & WN9 & 7 $\times$ 4.6 & 105 $\times$ 69       & 110 & 39 \\ 
            30 & O3.5If*/WN5 & 10.2 $\times$ 7.4 & 153 $\times$ 111 & 105 & (36) \\ 
            37 & WN3+OB & 13.8 $\times$ 12.8 & 207 $\times$ 192 & 137 & 43 \\ 
            41 & WN6h & 7.7 $\times$ 5.5 & 115.5 $\times$ 82.5      & 165 & ... \\ 
            43 & WC4+abs & 15.4 $\times$ 12.7 & 231 $\times$ 190.5  & 164 & 50 \\ 
            46 & WN3 & 7.6 $\times$ 3.8 & 114 $\times$ 57           & 175 & ((52)) \\ 
            47 & WC4+abs & 12 $\times$ 9.5 & 180 $\times$ 142.5     & 192 & 54 \\ 
            49 & WC4+O6III/V & 10 $\times$ 10 & 150 $\times$ 150    & 199 & 58 \\ 
            50 & WN4 & 10 $\times$ 10 & 150 $\times$ 150            & 199 & 58 \\ 
            52 & B3I+WN5 & 10 $\times$ 10 & 150 $\times$ 150        & 199 & 58 \\ 
            53 & WN3+OB & 14.4 $\times$ 13.5 & 216 $\times$ 202.5   & 208 & 62 \\ 
            61 & WN3+O7.5 & 14 $\times$ 11 & 210 $\times$ 165       & 221 & ((66)) \\ 
            65 & WC4+abs & 14 $\times$ 11 & 210 $\times$ 165        & 221 & 69 \\ 
            68 & WN3 & 11.5 $\times$ 6.5 & 172.5 $\times$ 97.5      & 229 & ((76)) \\ 
            73 & WC4 & 21 $\times$ 15 & 315 $\times$ 225            & 246 & 81 \\ 
            74 & B0I+WN & 21 $\times$ 15 & 315 $\times$ 225         & 246 & 81 \\ 
            77 & WN3+O & 21 $\times$ 15 & 315 $\times$ 225          & 246 & 81 \\ 
            80 & WN3+O8-9III & 21 $\times$ 15 & 315 $\times$ 225    & 246 & 87 \\ 
            99 & WC4(+OB) & 6.9 $\times$ 5.2 & 103.5 $\times$ 78    & 261 & 96 \\ 
            142 & WC4+abs & 7.5 $\times$ 6.3 & 112.5 $\times$ 94.5  & 269 & 104 \\ 
            143 & WN3/O3 & 12 $\times$ 8 & 180 $\times$ 120         & 284 & 103 \\ 
            144 & WN3+O7 & 7.5 $\times$ 6.3 & 112.5 $\times$ 94.5   & 269 & 104 \\ 
            145 & WC4+O & 7.5 $\times$ 6.3 & 112.5 $\times$ 94.5    & 269 & 104 \\ 
            151 & WN4 & 17 $\times$ 15 & 255 $\times$ 225           & 309 & 116 \\ 
            153 & WN11 & 17 $\times$ 15 & 255 $\times$ 225          & 308 & ((116)) \\ 
            154 & WN3 & 7.7 $\times$ 5.5 & 115.5 $\times$ 82.5      & 315 & ...  
        \enddata
       
  \end{deluxetable*}

  \begin{deluxetable*}{cccccc}
    \tablecaption{WR Stars in Superbubbles in 30 Dor}
    \tablehead{WR \#  & Spectral Type & \multicolumn{2}{c}{Dimensions} & Nebula & OB \\
                 &  & (arcmin) & (pc) & Name & \\
                (1) & (2) & (3) & (4) & (5) & (6)}
        \startdata
            82 & WC4        	& 7.5 $\times$ 5.2 & 112.5 $\times$ 78 & N157C & 90 \\
            83 & WN5h	        & 7.5 $\times$ 5.2 & 112.5 $\times$ 78 & N157C & 90 \\
            84 & O3.5If*/WN7	& 7.5 $\times$ 5.2 & 112.5 $\times$ 78 & N157C & 90 \\
            85 & WC4	        & 7.5 $\times$ 5.2 & 112.5 $\times$ 78 & N157C & 90 \\
            92 & WN7	        & 7.5 $\times$ 5.2 & 112.5 $\times$ 78 & N157C & 90 \\
            93 & WN4	        & 7.5 $\times$ 5.2 & 112.5 $\times$ 78 & N157C & 90 \\
            94 & WN7	        & 7.5 $\times$ 5.2 & 112.5 $\times$ 78 & N157C & 90 \\
            95 & WN5	        & 7.5 $\times$ 5.2 & 112.5 $\times$ 78 & N157C & 90 \\
            108 & B1I+WN3	    & 8 $\times$ 7     & 120 $\times$ 105  & Shell 3 & ... \\
            115 & WN6	        & 2.5 $\times$ 1.4 & 37.5 $\times$ 21  & Central Shell & 100 \\
            116 & O2.5If*/WN6	& 2.5 $\times$ 1.4 & 37.5 $\times$ 21  & Central Shell & 100 \\
            117 & WN6h      	& 2.5 $\times$ 1.4 & 37.5 $\times$ 21  & Central Shell & 100 \\
            121 & O2If*/WN5	    & 2.5 $\times$ 1.4 & 37.5 $\times$ 21  & Central Shell & 100 \\
            122 & O2If*     	& 2.5 $\times$ 1.4 & 37.5 $\times$ 21  & Central Shell & 100 \\
            123 & WN4.5h    	& 2.5 $\times$ 1.4 & 37.5 $\times$ 21  & Central Shell & 100 \\
            124 & WN5h      	& 2.5 $\times$ 1.4 & 37.5 $\times$ 21  & Central Shell & 100 \\
            125 & O2If*/WN5	    & 2.5 $\times$ 1.4 & 37.5 $\times$ 21  & Central Shell & 100 \\
            126 & WN5h      	& 2.5 $\times$ 1.4 & 37.5 $\times$ 21  & Central Shell & 100 \\
            127 & O2If*/O3If*/WN6	& 2.5 $\times$ 1.4 & 37.5 $\times$ 21  & Central Shell & 100 \\
            128 & WN4.5h    	& 2.5 $\times$ 1.4 & 37.5 $\times$ 21  & Central Shell & 100 \\
            129 & O2If*/WN5	    & 2.5 $\times$ 1.4 & 37.5 $\times$ 21  & Central Shell & 100 \\
            130 & O2If*/WN5	    & 2.5 $\times$ 1.4 & 37.5 $\times$ 21  & Central Shell & 100 \\
            131 & WC5	        & 2.5 $\times$ 1.4 & 37.5 $\times$ 21  & Central Shell & 100 \\
            132 & WN4.5h    	& 2.5 $\times$ 1.4 & 37.5 $\times$ 21  & Central Shell & 100 \\
            133 & WN4.5	        & 8 $\times$ 4 & 120 $\times$ 60  & Shell 5 & 100 \\
            134 & WN5/6+WN6/7	& 8 $\times$ 4 & 120 $\times$ 60  & Shell 5 & 100 \\
            136 & WN6+O3.5If*/WN7   & 2.5 $\times$ 1.4 & 37.5 $\times$ 21  & Central Shell & 100 
        \enddata
  \end{deluxetable*}

\begin{deluxetable*}{c|cc|cc|cc}
    \tablecaption{LMC WR Stars in LH OB Associations}
\tablehead{ &  \multicolumn{2}{c}{The LMC}    &   \multicolumn{2}{c}{The LMC $-$ 30 Dor} &  \multicolumn{2}{c}{30 Dor} \\
            Spec.\ Type   & \# WR  &\# WR in OB & \# WR & \# WR in OB &  \# WR & \# WR in OB  \\
             (1)   & (2)   & (3)        & (4)  & (5)        & (6)   & (7)}
        \startdata
            WN2-4  & 72 & 24 (33\%)  & 65   & 21 (32\%)  & ~7  & ~3 (43\%)  \\
            WN5-6  & 31 & 23 (74\%)  & ~8   & ~4 (50\%)  & 23  & 19 (83\%) \\
            WN7-L  & 23 & 11 (48\%)  & 16   & ~7 (44\%)  & ~7  & ~4 (57\%) \\
            WC4    & 21 & 16 (76\%)  & 15   & 13 (87\%)  & ~6  & ~3 (50\%)  \\ 
            WC5-6  & ~2 & ~2 ~(~...~)   & ~1   & ~1 ~(~...~)   & ~1  & ~1 ~(~...~)  \\
            WO3-4  & ~3 & ~2 ~(~...~)   & ~3   & ~2 ~(~...~)   & ~0  & ~0 ~(~...~)  \\
            Other  & ~4 & ~4 ~(~...~)   & ~2   & ~2 ~(~...~)   & ~2  & ~2 ~(~...~)  \\
            \hline 
            Total  &156 & 82 (53\%)  & 110  & 50 (45\%)  & 46  & 32 (70\%)
        \enddata
        \end{deluxetable*}

\begin{deluxetable*}{c|cccc|cccc|cc}
    \tablecaption{Spectral Types of WR Stars in Bubbles and Superbubbles}
    \tablehead{ &  \multicolumn{4}{c}{The LMC}    & \multicolumn{4}{c}{The LMC - 30 Dor} &        \multicolumn{2}{c}{30 Dor} \\
        Spectral & \# of WR in & In OB & \# of WR in & In OB &  \# of WR in & In OB & \# of WR in  & In OB & \# of WR in & In OB \\
              Type & Bubbles & Assc.& Superbubbles & Assc. & Bubble & Assc. & Superbubbles & Assc.  & Superbubbles & Assc. \\
            (1)    &(2) & (3) & (4) & (5) & (6) & (7) & (8) & (9) &(10)&(11)}
        \startdata
         WN2-4  & 12 (17\%)   & 0  & 24 (33\%)  & 13   & 11 (17\%) & 0  & 20 (31\%)  & 10  & ~4 (57\%)  & 3 \\
         WN5-6  & ~2 (~6\%)   & 1  & 20 (65\%)  & 18   & 2  (25\%) & 1  & 3  (38\%)  & 1   & 17 (74\%) & 17 \\
         WN7-L  & ~2 (~9\%)   & 0  & ~7 (30\%)  & 5    & 2  (13\%) & 0  & 3  (19\%) & 1   & 4 ~(~...~) & 4  \\
         WC4    & ~2 (10\%)   & 2  & 11 (52\%)  & 11   & 2  (13\%) & 2  & 9  (60\%) & 9   & 2 ~(~...~) & 2  \\ 
         WC5-6  & ~0 ~(~...~) & 0  & ~1 ~(~...~) & 1    & 0~(~...~) & 0   & 0 ~(~...~)  & 0   & 1 ~(~...~)  & 1 \\
         WO3-4  & ~0 ~(~...~) & 0  & ~0 ~(~...~) & 0    & 0~(~...~) & 0   & 0 ~(~...~)   & 0   & 0 ~(~...~)  & 0\\
         Other  & ~0 ~(~...~) & 1  & ~2 ~(~...~) & 2    & 0~(~...~) & 0   & 1 ~(~...~)   & 1   & 1 ~(~...~)  & 1 \\
            \hline 
         Total  & 18 (12\%)   & 4  & 65 (42\%)   & 50   & 17 (15\%) & 3   & 36 (33\%)  & 22  & 29 (63\%) & 28
        \enddata
\end{deluxetable*}

\begin{deluxetable*}{c|ccc|ccc|ccc}
    \tablecaption{\ion{H}{2} Region Morphology Classes of Different Types of WR Stars}
    \tablehead{ &  \multicolumn{3}{c}{The LMC}    & \multicolumn{3}{c}{LMC--30\,Dor} &        \multicolumn{3}{c}{30 Dor} \\
        Spectral & \multicolumn{3}{c}{\ion{H}{2} Classes} & \multicolumn{3}{c}{\ion{H}{2} Classes} & \multicolumn{3}{c}{\ion{H}{2} Classes} \\
              Type & 1   & 2   &  3  &  1  &  2  &  3   & 1  &   2 & 3}  
        \startdata
         WN2-4     & ~3  & 26  & 43  & ~3  & 21  & 41  & ~0  & ~5  & ~2 \\
         WN5-6     & ~2  & 23  & ~6  & ~0  & ~4  & ~4  & ~2  & 19  & ~2 \\
         WN7-L     & ~5  & ~6  & 12  & ~2  & ~2  & 12  & ~3  & ~4  & ~0 \\
         WC4       & ~2  & 14  & ~5  & ~1  & 11  & ~3  & ~1  & ~3  & ~2 \\ 
         WC5-6     & ~0  & ~1  & ~1  & ~0  & ~0  & ~1  & ~0  & ~1  & ~0  \\
         WO3-4     & ~0  & ~0  & ~3  & ~0  & ~0  & ~3  & ~0  & ~0  & ~0 \\
         Other     & ~1  & ~3  & ~0  & ~0  & ~2  & ~0  & ~1  & ~1  & ~0 \\
         \hline 
         Total     & 13  & 73  & 70  & ~6  & 40  &  64 & ~7  & 33  & ~6
        \enddata
\end{deluxetable*}

\clearpage
\appendix

\hangindent=0in
In this appendix, the nomenclature for OB associations, \hi and \hii features are abbreviated as the following:

\smallskip

{\bf LHn} -- OB association from \citet{LH1970},

{\bf Nn} -- LH$\alpha$ 120-Nn \hii region from \citet{Henize1956},

{\bf DEM\,Ln} -- LMC \hii region from \citet{DEM1976},

{\bf \hii SGS LMC-n} -- ionized SGS identified from \ha

\indent
~~~images by \citet{Meaburn1980},

{\bf \hi GS\,n} -- \hi giant shell from \citet{Kim1999}, and

{\bf \hi SGS\,n} -- \hi SGS from \citet{Kim1999}.

\bigskip
{\bf WR1} is in a field without any nebulosity in the vicinity or any identifiable large shell structure. 

{\bf WR2} is surrounded by a small, incomplete shell with dimensions 0\farcm 4 $\times$ 0\farcm3 on the northern rim of a larger 3\farcm2 $\times$ 2\farcm2 shell, which is likely a superbubble blown collectively by WR2 and other massive stars. The small shell was first reported by \citet{Pakull1991} to be highly ionized, and its abundance was observed by \citet{Garnett1994}. It was later shown that both the small and large shells exhibit \ion{He}{2} $\lambda$4686 emission \citep{Naze2003}. This small shell has an expansion velocity of $\sim$16 \kms \citep{Chu1999}. On a larger scale, these features are projected on the northwestern rim of \hii SGS LMC-7 \citep{Meaburn1980}. There is a known supernova remnant, MCELS J0449-6921 \citep{Maggi2016}, centered $\sim$1\farcm5 to the northwest of WR2, and it is best seen in the \sii image. WR2 is more than 100 pc from LH1, thus not likely a member of this OB association.

{\bf WR3} is located in a field with a network of faint \ha\ filaments without obvious shell structure. In the 8 $\mu$m and 24 $\mu$m images, there is a long, thin filament extending from the eastern vicinity to 1\farcm6 north of the star; however, it is uncertain whether this filament is physically associated with WR3. The background \hi gas shows two radial velocity components at 282 and 295 km~s$^{-1}$, which may indicate a large-scale expanding structure, but no \hi giant shells have been identified. The relationship between the gas kinematics and the WR star is unclear.

{\bf WR4} is surrounded by a 0\farcm6 shell on the eastern rim of a 4\farcm3 $\times$ 4\arcmin\ shell extending from the luminous \hii region DEM\,L10 around the OB association LH2. The small shell is best seen in the 24 $\mu$m image. WR4 is inside \hi GS\,9, as corroborated by the split velocity components in the \ion{H}{1}. The WR star is projected at $\sim$80 pc from LH2, thus not likely a member of this OB association. Details of this object have been reported by \citet{Gvaramadze2014}.

{\bf WR5} is in a complex network of faint \ha\ filaments located on the northern rim of \hii SGS LMC-7, and some filaments appear to form a large shell measuring 18\arcmin\ $\times$ 12\arcmin\ with its major axis oriented along the NE-SW direction. Neither the shell nature nor the physical association between the star and the shell is certain. WR5 is also projected inside \hi GS\,9. This \hi shell is more extended than the \ha\ shell in the southeast direction. The star is $\sim$90 pc from LH5, thus not likely a member of this OB association.

{\bf WR6} is superposed on diffuse \ha\ emission in the outskirts of the luminous \hii region DEM\,L22 and associated with the OB association LH5. These features are all on the northern rim of \hii SGS LMC-7. The WR star is projected inside \hi GS\,15, which is corroborated by the \hi position-velocity plot. Note, however, that \hii SGS LMC-7 is much more extended than \hi GS\,15.

{\bf WR7} is superposed on diffuse \ha\ emission and projected at $\sim$80 pc from LH5. The 8 $\mu$m and 24 $\mu$m images show a 4\arcmin\ long filamentary arc structure to the west of the star, roughly following the surface of the \hi cloud. The infrared arc structure and the apparent \hi cavity around the WR star suggest a wind-ISM interaction. The WR star is projected in the interior of \hi GS\,15 and on the northern rim of \hii SGS LMC-7.

{\bf WR8} is projected within a faint, filamentary shell-structure with dimensions 11\arcmin\ $\times$ 7\farcm6. The \oiii$\lambda$5007/\ha\ ratio of this shell is higher than other \hii regions, indicating a higher excitation. As WR8 is an early type WN3 star with a high effective temperature, it is likely that WR8 has photoionized this nebula.  

{\bf WR9} is projected in an incomplete, filamentary shell-like structure with dimensions 7\farcm6 $\times$ 6\farcm6 and associated with the OB association LH8. The WR star is also on the western rim of \hi GS\,18 and inside \hii SGS LMC-7. The \hi position-velocity plots show split velocity components from the expansion of \hi GS\,18.

{\bf WR10} is located in the northern outskirts of the \hii complex N11.  The overall morphology of this region suggests an outflow from the prominent central superbubble around LH9, and the \hi position-velocity plots show velocity splits indicating expansion.  In the \ha\ image, WR10 is surrounded by a 4$'$ $\times$ 2$'$ arc in the north, resembling a half shell, and a straight filament to its west; however, the [\ion{O}{3}] image shows that the straight filament is of lower excitation and thus extraneous.  The WR star may be partially responsible for the half-shell structure. WR10 is more  than 100 pc away from the OB associations LH9 and LH10, thus, it cannot be a member of either.

{\bf WR11} is in the prominent 12\arcmin\ $\times$ 8\arcmin\ superbubble in the \hii complex N11 and a member of the central OB association LH9. The superbubble is inside \hi GS\,16. The expansion of the superbubble is clearly seen in the \hi position-velocity plots.

{\bf WR12} is in a 3\arcmin\ $\times$ 2\arcmin\ shell in the \hii region DEM\,L39 around the OB association LH12, of which the WR star is a member. This OB association and its \hii region are on the northwestern rim of \hi SGS\,2 and \hii SGS LMC-6. The bubble of WR12 was first identified by \citet{Chu1980} and has been observed to have an average expansion velocity of 42 \kms \citep{Chu1983}.

{\bf WR13} is located in \hi GS\,18 and \hii SGS LMC-7. The 8 $\mu$m and 24 $\mu$m images show an arc around the star, which measure $\sim$3\farcm5 across. The northeast end of this arc is connected to more emission features. It is difficult to discern whether these features are physically associated with WR13.

{\bf WR14} is a known runaway star with a radial velocity of $\leq$130 \kms relative to the ambient medium \citep{Gvaramadze2010}. There is a 2\farcm2 $\times$ 1\farcm4 bow-shock-like feature northeast of the star, DEM\,L40; however, the WR star is not at its center of curvature. The \hi gas shows two velocity components, although the WR star is not in any catalogued \hi shells. 

{\bf WR15} is surrounded by some complex filamentary features without apparent shell morphology. WR15 is located on the southeastern rim of the N11 complex. The WR star is not in any catalogued OB associations in N11. 

{\bf WR16} does not show any obvious \ha\ emission in its vicinity. The WR star is located inside \hi SGS\,2 and \hii SGS LMC-6. Note that the coordinates of this star in Table 1 have been corrected from \citet{Neugent2018}.

{\bf WR17} is in a small 2\arcmin\ $\times$ 2\arcmin\ shell \citep{Dopita1994} inside a larger 10\farcm2 $\times$ 8\farcm4 superbubble in DEM\,L45 \citep{Chu1980}.  The small shell is best seen in the \oiii line because the spectral type of the WR star is WN3, thus, its photoionized gas has high excitation and high \oiii$\lambda$5007/\ha\ ratio.

{\bf WR18} has no obvious nebulosity in its vicinity; however, on a larger scale, it is surrounded by a faint shell-like structure with dimensions 20\arcmin\ $\times$ 20\arcmin. The northern part of the shell is brighter and coincides with \hi GS\,22.  The relationship between the optical shell and the \hi shell is not clear as the former is twice as extended as the latter.  WR18 is projected outside \hi GS\,22.

{\bf WR19} has a small, incomplete elliptical shell surrounding the star with dimensions 0\farcm7 $\times$ 0\farcm3 \citep{Dopita1994}. This shell's nitrogen abundance indicates enrichment by stellar ejecta \citep{Garnett1994,Stock2011}, and its observed average expansion velocity is 80 \kms \citep{Chu1999}. The WR star is also on the northwest outskirts of the \hii region DEM\,L5.

{\bf WR20} is in the diffuse emission region DEM\,L68 adjacent to the bright \hii region DEM\,L67. The \oiii image shows a diffuse large ring with a much higher \oiii$\lambda$5007/\ha\ ratio than the bright \hii region DEM\,L67. The ring measures 7\farcm7 $\times$ 4\farcm3 and might be associated with the WR star. The star is also projected on the periphery of \hii SGS LMC-8. The \hi position-velocity plots show velocity splits, but no \hi shell was cataloged.

{\bf WR21} is inside a 3\farcm0 $\times$ 2\farcm8 shell-like structure in the \hii region DEM\,L66. The WR star is not in known OB associations. WR21 is also projected on the northern rim of \hi SGS\,5. The \hi gas shows two velocity components possibly associated with \hi SGS\,5.  

{\bf WR22} has no nearby nebulosity but is located in \hii SGS LMC-8. The WR star is over 100 pc away from LH18 and LH26, which is too distant for a membership.

{\bf WR23} is surrounded by some faint diffuse and filamentary \ha\ emission features but with no recognizable shell structures. The WR star is within 50 pc to the rim of LH26 but over 100 pc to its center, thus, it is unlikely to be a member. WR23 is also projected in \hii SGS LMC-8.

{\bf WR24} is located in a blister-like structure of dimensions 4\farcm4 $\times$ 3\arcmin\ extending from the bright \hii region DEM\,L86 to the west. There are fainter filaments extending to the north and south of this blister structure, but the relationship between them is not clear. DEM\,L86 is photoionized by the OB association LH31, of which WR24 is a member. The WR star is near the eastern interior of \hi SGS\,5. The \hi gas shows two velocity components possibly associated with the expansion of \hi SGS\,5.

{\bf WR25} is at the base of an east-west elongated shell, measuring 2\farcm8 $\times$ 1\farcm4 and extending from the bright \hii region DEM\,L86 to the east \citep{Dopita1994}. DEM\,L86 is photoionized by the OB association LH31, of which WR25 is a member. The complex is on the eastern rim of \hi SGS\,5. The \hi gas shows two velocity components possibly associated with the expansion of SGS\,5.

{\bf WR26} is on the bright southern rim of the 10\farcm2 $\times$ 7\farcm4 shell structure of DEM\,L105, which is centered on the OB association LH36. However, WR26 is a member of a different OB association, LH37, on the southern rim of DEM\,L105. There are no sharp filaments near WR26 indicating stellar wind interaction. The whole structure is inside \hi GS\,44.

{\bf WR27} is a member of the OB association LH39 and inside the incomplete shell structure DEM\,L110 with dimensions 7\arcmin\ $\times$ 4\farcm6. It is likely that the OB association is responsible for shaping the shell structure. Interestingly, the 8 $\mu$m image shows that WR27 is near the center of an apparent cavity.  DEM\,L110 is inside \hi GS\,45. 

{\bf WR28} is superposed on some diffuse \ha\ emission without any filamentary or shell morphology between the shell \hii regions DEM\,L105 and DEM\,L106, which are centered on the OB associations LH36 and LH38, respectively. At a distance of $\sim$30--35 pc from each OB association, it is not clear whether the WR star is a field star or a runaway star originating from one of these OB associations. The WR star and the neighboring \hii regions DEM\,L105 and DEM\,L106 are located inside \hi GS\,44.

{\bf WR29} is inside a 2\farcm8 $\times$ 2\farcm6 shell structure extending from the bright \hii region DEM\,L108. The shell is better seen in the 8 $\mu$m and 24 $\mu$m images. The WR star is $\sim$10 pc from the edge the OB association LH35 and within 50 pc from the center. WR29 is projected within \hi GS\,46.

{\bf WR30} is projected near the eastern interior of the 10\farcm2 $\times$ 7\farcm4 shell \hii region DEM\,L105 and is located $\sim$25 pc outside the OB association LH36, which is at the center of DEM\,L105. The WR star and DEM\,L105 are inside \hi GS\,44.

{\bf WR31} is in the diffuse \hii region DEM\,L119 with some filaments on the bright southern part of the \hii region. The WR star is projected on the southwest rim of \hi GS\,47. 

{\bf WR32} is superposed on some diffuse emission to the west of DEM\,L132a. There are some random filaments nearby but no organized shell structure. The WR star is projected on \hi GS\,51. The \hi gas shows two velocity components associated with the expansion of \hi GS\,51.

{\bf WR33 and WR34} is in a tight cluster in the OB association LH41 and located inside the \hii region DEM\,L132a. The cluster is in \hi GS\,54.

{\bf WR35} is on the edge of the \hii region DEM\,L132b and is a member of the OB association LH41. Due to the large number of massive stars in the vicinity, multiple clusters in LH41, and complexity in the nebula structure, it is not possible to identify specific features associated with the WR star. No small bubbles are seen around the star. The star is projected in \hi GS\,54.

{\bf WR36} is a member of the OB association LH42 and is on the northwest rim of a small 1\farcm4 $\times$ 1\farcm1 shell structure, though they are likely not associated. The 8 $\mu$m image shows a 2\farcm6 $\times$ 2$'$ cavity around the WR star, and the \hi position velocity plots also show a slowly expanding shell structure around the WR star. The WR star could be responsible for excavating this cavity. The bright \hii region DEM\,L143 is associated with the OB association LH42.

{\bf WR37} is located in the OB association LH43 and inside the large 13\farcm8 $\times$ 12\farcm8 superbubble DEM\,L137 \citep{Chu1980}. The superbubble expansion velocity was observed to be 15--20 km s$^{-1}$ \citep{Chu1982b}.  The WR star is also projected in \hi SGS\,6.

{\bf WR38, WR 39} are near some nebulosities and are associated with the OB association LH45. These two WR stars are inside \hii SGS LMC-5 with complex filamentary structure.  No specific structures can be identified to be specifically associated with these WR stars.  \hii SGS LMC-5 is associated with \hi SGS\,7.

{\bf WR40} is located inside the OB association LH47 on the southwest rim of the superbubble in DEM\,L152. The WR star is also in the base of a blowout-like structure extending 1\farcm8 long and 1\farcm2 wide.

{\bf WR41} is in the large elliptical shell DEM\,L165 of dimensions 8\farcm1 $\times$ 5\farcm5 with an opening on the southeast quadrant while the south rim of the shell extends from the base of the major axis to the northeast for about 6$'$ \citep{Chu1980}. The expansion velocity of this shell has been determined to be less than 10 \kms \citep{Chu1982b}.  The \hi position-velocity plot indicates an expansion with an expansion velocity of $\sim$12 \kms. The WR star is not in any known OB associations and is located on the south rim of \hi GS\,61. 

{\bf WR42} is in the bright diffuse \hii region DEM\,L160, which has nonuniform surface brightness with dust lanes and emission filaments but does not have any discernable shell structure around the star. The WR star is a member of LH49.

{\bf WR43} is inside the 15\farcm4 $\times$ 12\farcm7 superbubble DEM\,L164. The superbubble is around the OB association LH50, of which WR43 is a member. DEM\,L164 is on the south rim of \hii SGS LMC-9 and inside \hi GS\,61. The largest line splitting of the \hi position-velocity plot shows an expansion of \hi GS\,61 of up to 20 \kms. 

{\bf WR44} is projected inside the \hii SGS LMC-5. Some faint \ha\ filaments are seen within the central cavity of LMC-5 and together, these filaments are called DEM\,L154. The filament to the south of the star is curved away from the star, thus, it is unlikely that the WR star is responsible for shaping this feature. The WR star is also inside \hi SGS\,7.

{\bf WR45} is in the \hii region DEM\,L174, which shows striations in its morphology \citep{Chu1980}. Although there is an apparent cavity near the star and some curved filaments suggesting wind-ISM interaction, the overall \hii morphology does not indicate any shell structure. 
The internal motion does not indicate an expansion \citep{Chu1982b}. The WR star is projected outside the northwest rim of \hii SGS LMC-3. 

{\bf WR46} is in a small emission nebula with numerous dust lanes to the north of the star. This small nebula is interior to the northern rim of a 7\farcm6 $\times$ 3\farcm8 shell structure. WR46 is $\sim$80 pc from the OB association LH52 thus likely not a member of this OB association. The WR star is also projected on the eastern rim of \hii SGS LMC-5 and in \hi SGS\,7.

{\bf WR47} is inside the 12 $\times$ 9\farcm5 superbubble DEM\,L192 and a member of the OB association LH54. DEM\,L192 is projected on the southwest rim of \hii SGS LMC-4 and \hi SGS\,11. \citet{Stock2011} reported a small nebula around this star and that its elemental abundances are similar to those of LMC \hii regions. However, \emph{Hubble Space Telescope} images have resolved this small nebula into bright-rimmed dust globules, and \emph{Spitzer Space Telescope} has revealed embedded star formation \citep{Chu2005}; thus, this nebula has no association with the WR star. These bright-rimmed dust globules can be seen in the MCELS2 \ha\ image, and the young stellar objects in the dust globules can be seen in the 8 $\mu$m and the 24 $\mu$m images. The surface of the globules might be photoionized by WR47, but this nebula is not shaped by the WR star and there is no evidence of wind-ISM interaction. 

{\bf WR48} is inside a 3\farcm8 $\times$ 2\farcm9 shell structure in the southeast lobe of DEM\,L175a while the northwestern lobe is a supernova remnant. Overall, the structure is shaped like a double-lobed nebula and is located in the interaction zone between \hii SGS LMC-4 and \hii SGS LMC-5 and the interaction zone between \hi SGS\,7 and SGS\,11. The shell's large size, 50 pc across, makes it uncertain whether it is a bubble blown by WR48.

{\bf WR49 and WR50} are members of the OB association LH58 in the 10$'$ $\times$ 10$'$ superbubble DEM\,L199, which is on the west rim of \hii SGS LMC-3. The WR star is also in \hi SGS\,12.

{\bf WR51} is projected along a curved filament of the \hii region DEM\,L198, which is inside \hii SGS LMC-3. The WR star is also in \hi SGS\,12.

{\bf WR52} is member of the OB association LH58 in the 10$'$ $\times$ 10$'$ superbubble DEM\,L199, which is on the west rim of \hii SGS LMC-3. A continuum-subtracted [\ion{O}{3}]/\ha\ ratio map reveals a small 0\farcm8 $\times$ 0\farcm8 shell around the star \citep{Oey2000}. While this small shell is prominent in the [\ion{O}{3}]/\ha\ map, only partial northwestern and southeastern rims of this shell can be seen in the MCELS2 \ha\ and MCELS1 \oiii images. This small bubble has also been noted by \citet{Dopita1994}. The WR star is also in \hi SGS\,12.

{\bf WR53} is in the OB association LH62 and inside the superbubble DEM\,L208, measuring 14\farcm4 $\times$ 13\farcm5 and has multiple rings most prominent in the \ha\ image \citep{Chu1980}. DEM\,L208 is on the northeastern rim of \hii SGS LMC-9.

{\bf WR54} is superposed on the faint, diffuse \hii region DEM\,L210 with a filamentary shell-like structure, measuring 3\arcmin\ $\times$ 2\farcm4, extending from the star to the south. DEM\,L210 is projected inside in \hii SGS LMC-3 and \hi SGS\,12.

{\bf WR55} is in the faint diffuse emission region DEM\,L210 and is a member of the OB association LH61. DEM\,L210 is projected within \hii SGS LMC-3 and \hi SGS\,12.

{\bf WR56} is in the faint diffuse emission region DEM\,L210 and is $\sim$60 pc from LH61 thus not a member of this OB association. DEM\,L210 is projected within \hii SGS LMC-3 and \hi SGS\,12.

{\bf WR57} is in the faint diffuse emission region DEM\,L210, which is projected within \hii SGS LMC-3 and \hi SGS\,12.

{\bf WR58} is a member of the OB association LH64 and projected on the northern interior of \hii SGS LMC-3 and \hi SGS\,12.

{\bf WR59} is projected on the southern interior of \hii SGS LMC-3 and \hi SGS\,12.

{\bf WR60} is located near some diffuse nebulosity. The WR star is inside \hii SGS LMC-3 and \hi SGS\,12. 

{\bf WR61} is surrounded by an apparent bow-shock like feature, measuring 1\farcm4 $\times$ 1\farcm4, which has been known to emit \ion{He}{2}$\lambda$4686 emissions \citep{Niemela1991}. However, the MCELS2 \ha\ and MCELS1 [\ion{O}{3}] images show that the ``bow-shock'' may not be a coherent structure. The WR star is $\sim$70 pc from LH66 and more than 100 pc from LH69, although it is projected within the 14$''$ $\times$ 11$''$ superbubble DEM\,L221. The WR star is also projected in \hii SGS LMC-9 and \hi GS\,70.  The \hi position-velocity plots show split components originating from the expansion of \hi GS\,70. 

{\bf WR62} is in a cavity of \hii SGS LMC-3 and \hi SGS\,12 with no nearby nebulosity.

{\bf WR63} is $\sim$70 pc from the OB association LH64 and is projected on the northern interior of \hii SGS LMC-3 and \hi SGS\,12.

{\bf WR64} is on the southern interior of \hii SGS LMC-4 without any obvious nebulosity nearby. The WR star is also projected within \hi SGS\,11, whose expansion is corroborated by the two velocity components in the \hi position-velocity plots.

{\bf WR65} is near some filamentary structure within the superbubble DEM\,L221 and on the southwest rim of \hii SGS LMC-9. This WR star is a member of the OB association LH69, which is responsible for shaping the 14$'$ $\times$ 11$'$ superbubble DEM\,L221 and the surrounding \hi GS\,70.

{\bf WR66} is superposed on a filament on the eastern rim of \hii SGS LMC-3 and $\ge$100 pc from the OB associations LH74 and LH67 thus not likely a member of either.

{\bf WR67} is near some filaments on the eastern rim of \hii SGS LMC-3 and $\ge$100 pc from the OB associations LH74 and LH67 thus not likely a member of either. It is interesting to note that WR67 is a very late type WN star (WN11); the bright 24 $\mu$m emission indicates a very dense stellar wind.  There is a small emission nebula at 15$''$ northeast of the WR star, and may consist of material ejected by the star. Spectroscopic observations of the nebular abundance are needed to determine its nature.

{\bf WR68} is $\sim$60 pc from LH76, thus likely not a member. The WR star is also projected on the western rim of the 11\farcm5 $\times$ 6\farcm5 superbubble DEM\,L229, which is blown by LH76 and on the southern rim of the \hii SGS LMC-4. 

{\bf WR69} is not superposed on any detectable nebulosity but is projected near a faint filament on the southern interior of \hii SGS LMC-4 and within \hi SGS\,11. The WR star is also $\sim$90 pc from LH70 thus likely not a member.

{\bf WR70} is located in a dusty, complex environment of the \hii region DEM\,L227 with a large arc northeast of the star, but no shell structure can be claimed to be formed by this WR star. DEM\,L227 is located on the northern rim of \hii SGS LMC-3, associated with \hi SGS\,12.

{\bf WR71} is $\sim$75 pc from the OB association LH76, thus likely not a member. The WR star is inside the small 1\farcm5 $\times$ 1\farcm2 shell DEM\,L231 \citep{Chu1980}, with an expansion velocity no greater than 18 $\pm$ 1 \kms \citep{Chu1983}. 

{\bf WR72} is in a diffuse emission field of DEM\,L224, which contains a collection of random filaments without any shell-like structure. This WR star is not in any known OB association.

{\bf WR73, WR74, and WR77} are members of the OB association LH81, which is projected inside the 21$'$ $\times$ 15$'$ superbubble DEM\,L246 within \hi GS\,73. The \hi position-velocity plots show the expansion of \hi GS\,73.

{\bf WR75} is inside a 0\farcm7 $\times$ 0\farcm2 bow shock-like structure embedded in the brightest parts of the filamentary conglomeration of DEM\,L239. The bow shock structure is best seen in the  MCELS2 \ha\ image. DEM\,L239 is on the northern rim of \hii SGS LMC-4 and \hi SGS\,11.

{\bf WR76} is in a small 1\farcm5 $\times$ 1\farcm1 bubble in DEM\,L240 on the southeastern edge of \hii SGS LMC-4 and \hi SGS\,11. This shell was first observed by \citet{Chu1980}, and \citet{Cowley1984} reported the high radial velocity of the star, $\sim$470  km s$^{-1}$, with respect to the LMC velocity, $\sim$270 km s$^{-1}$.

{\bf WR78} is superposed on the filamentary conglomeration DEM\,L263. There are no obvious wind interaction features that can be associated with this star.

{\bf WR79} is in a 0\farcm5 $\times$ 0\farcm3 ring associated with a small 0\farcm8 diameter \hii region embedded in a diffuse field within the conglomeration of DEM\,L263. This ring structure, identified as an oval ring nebula by \citet{Dopita1994} and studied by \citet{Stock2010}, is best seen in the MCELS2 \ha\ image. Comparisons between the \ha\ and [\ion{O}{3}] images indicate that this ring nebula has a higher [\ion{O}{3}]/\ha\ emission line ratio than the neighboring \hii region; this high excitation is expected from WR79's WN3 spectral type. WR79 is $\sim$90 pc from the rim of LH89 thus likely not a member. 

{\bf WR80} is a member of the OB association LH87. The star is projected inside the 21$'$ $\times$ 15$'$ superbubble DEM\,L246, which is inside \hi GS\,73. To the east of the WR star a filamentary shell structure can be seen along the northeast rim of the superbubble DEM\,L246.

{\bf WR81} is projected on the eastern interior of \hii SGS LMC-4 and \hi SGS\,11. The WR star is superposed on diffuse emission with no apparent stellar wind interaction features. 

{\bf WR82, WR83, WR84, and WR85} are members of LH90 and are projected along the western rim of the 7\farcm5 $\times$ 5\farcm2 superbubble 30 Dor C. The \hi giant shell associated with this superbubble is \hi GS\,75.  The \hi position-velocity plots show that \hi GS75 is more extended than the ionized superbubble. 

{\bf WR86, WR87, WR89, and WR91} are superposed on faint, diffuse emission between 30 Dor and \hii SGS LMC-3 and are members of LH89.

{\bf WR88} is superposed on faint, diffuse emission between 30 Dor and \hii SGS LMC-3 and is a member of LH85.

{\bf WR90} is projected on the eastern interior of \hii SGS LMC-4 and \hi SGS\,11. The WR star is superposed on vey faint diffuse emission with no apparent stellar wind interaction features. 

{\bf WR92, WR93, WR94, and WR95} are members of LH90 near the center of superbubble 30 Dor C, measuring 7\farcm5 $\times$ 5\farcm2. The \hi giant shell associated with this superbubble is \hi GS\,75, and its expansion velocity assessed from the \hi velocity splits is 12--15 km s$^{-1}$.

{\bf WR96} is a member of LH88 and is projected in a supernova remnant in DEM\,L241.

{\bf WR97} has an incomplete shell of radius 0\farcm75 that can be seen in the \ha\ image but shows the highest contrast against the background in the [\ion{O}{3}] image. On a larger scale, there appears to be a 6\farcm6 $\times$ 2\farcm7 shell that can be seen in the \ha\ and [\ion{O}{3}], and the MCELS2 \ha\ image reveals sharp filaments along the shell rim. The filaments on the eastern side of the shell structure appear to be connected with other sharp filaments on the southeast rim of 30 Dor C, forming a large arc structure and making the apparent superbubble structure around WR97 highly uncertain. The WR star is located on the northeastern outskirts of 30 Dor C and is $\sim$45 pc from the OB association LH90 thus likely not a member.

{\bf WR98} is a member of LH94 and located in an empty field between filamentary \ha\ emission features that do not seem to be associated with the WR star or LH94. The \hi position-velocity plots show multiple velocity components from the ISM.

{\bf WR99} is superposed on a band of \ha\ emission without any wind interaction features and is a member of LH96. On a larger scale, the WR star is surrounded by a 6\farcm9 $\times$ 5\farcm2 filamentary shell-like structure south of 30 Dor C.

{\bf WR100} is just outside the eastern rim of the superbubble 30 Dor C with no associated wind interaction features. The WR star is $\sim$60 pc from LH90 thus likely not a member of this OB association.

{\bf WR101} is in a diffuse emission region on the western edge of 30 Dor. No wind interaction features are seen in the vicinity of the star. 

{\bf WR102} is superposed on some diffuse field emission in DEM\,L261 and south of 30 Dor B without any wind interaction features. The WR star is a member of LH97.

{\bf WR103 and WR104} are in a diffuse emission region on the western edge of 30 Dor and $\sim$25 and $\sim$33 pc from the center of LH99, respectively. No wind interaction features are seen in the vicinity of the stars. The \hi position-velocity plots show very complex motions in the ISM.

{\bf WR105} is superposed on diffuse emission to the south of 30 Dor B without any wind interaction features and is $\sim$60 pc from the center of LH99.  The \hi position-velocity plots show at least four velocity components, indicating a very complex interstellar environment.

{\bf WR106} is in a small cluster with diffuse field emission on the northern edge of 30 Dor B and a member of LH99. No wind interaction features are seen in the vicinity of the star. 

{\bf WR107} is in an apparent cavity of dimensions 4\farcm3 $\times$ 1\farcm6. There is enhanced emission around the cavity, but it does not have sharp features indicating wind-ISM interaction. It is not clear whether the WR star is responsible for this structure. The \hi position-velocity plots show complex motions in the ISM. The WR star is located in DEM\,L261 to the south of 30 Dor B and is a member of LH97.

{\bf WR108} is superposed on diffuse emission with some unorganized filaments in Shell 3 of 30 Dor \citep{Wang1991}.  It is $\sim$70 pc from LH100 and $\sim$75 pc from LH99 and is not likely a member of either OB association. 

{\bf WR109} is superposed on diffuse emission on the northeastern part of 30 Dor B without any wind interaction features. It is a member of LH99.

{\bf WR110 and WR111} are superposed near some faint diffuse emission in DEM\,L269 without any wind-ISM interaction features. It is a member of LH101.

{\bf WR112} is 20 pc northwest of the R136 cluster inside 30 Dor. The WR star is outside the western rim of a 1\farcm1 $\times$ 0\farcm9 shell.

{\bf WR113 and WR114} are projected at 12--15 pc southwest of the R136 cluster in 30 Dor. These WR stars are outside the 2\farcm5 $\times$ 1\farcm4 shell around R136. They are superposed on very bright nebulosities, though no wind-ISM interaction features can be seen. 

{\bf WR115} is on the outskirts of the R136 cluster, LH100, in the center of 30 Dor and is interior to the western end of a 2\farcm5 $\times$ 1\farcm4 shell.

{\bf WR116, WR117, WR121, WR122, WR123, WR124, WR125, WR126, WR127, WR128, WR129, WR130, WR131, and WR132} are in the R136 cluster, or LH100, in the center of 30 Dor. This cluster is interior to the western end of a 2\farcm5 $\times$ 1\farcm4 shell.
30 Dor is in \hi GS\,78.  The \hi column densities are so high at the core of 30 Dor 
that self-absorption occurs, causing the apparent depression in surface brightness. 

{\bf WR118, WR119, WR120} are at $\sim$12 pc north of the R136 cluster in 30 Dor, but are still within LH100. They are outside the northern edge of the 2\farcm5 $\times$ 1\farcm4 shell around the R136 cluster and interior to the southern rim of a 1\farcm1 $\times$ 0\farcm9 shell. These three WR stars are projected in the boundary between the two shells and are likely associated with the latter shell.

{\bf WR133, WR134} are 80 and 60 pc, respectively, north of the R136 cluster, LH100, in 30 Dor. Both are inside Shell 5 of 30 Dor \citep{Wang1991}, measuring 8$'$ $\times$ 4$'$. 

{\bf WR135} is in 30 Dor but is $\sim$30 pc from the R136 cluster. The WR star is superposed on the edge of a dark cloud. No wind-ISM interaction features can be seen.

{\bf WR136} is not in the R136 cluster in the center of 30 Dor but is interior to the eastern end of the same 2\farcm5 $\times$ 1\farcm4 shell around the R136 cluster.

{\bf WR137} is superposed on bright diffuse emission in DEM\,L269 with no obvious wind interaction features associated with the star. The WR star is a member of LH101.

{\bf WR138} is superposed on bright emission of 30 Dor and is $\sim$45 pc from the R136 cluster, LH100. No wind-ISM interaction features can be identified close to the star, and the environment is too complex to unambiguously make any associations.

{\bf WR139} is projected just outside the eastern rim of Shell 5 of 30 Dor \citep{Wang1991}.  It is $\sim$75 pc from LH100 and thus likely not a member.

{\bf WR140} is in the faint diffuse outskirts of 30 Dor, about 330 pc from the R136 cluster.  There are some broad filamentary structures in the field, but none are curved around the WR star to indicate wind-ISM interaction.

{\bf WR141} is in the bright \hii region on the southwest rim of the superbubble in DEM\,L284, which sits in the ridge of very active star formation on the west rim of \hii SGS LMC-2. The surrounding of WR141 is quite dusty, and the apparent shell morphology in the [\ion{O}{3}] image is caused by embedded dust lanes. This environment is too complex to identify physical structures unambiguously. The WR star is a member of LH103.

{\bf WR142, WR144, WR145} are in the central cavity of the 7\farcm5 $\times$ 6\farcm3 superbubble of DEM\,L269, which sits in the ridge of very active star formation on 
the west rim of \hii SGS LMC-2, corresponding to \hi SGS\,19. The WR star is a member of LH104 that is responsible for the superbubble in DEM\,L269.

{\bf WR143} is projected within the 12$'$ $\times$ 8$'$ superbubble of DEM\,L284, which sits in the ridge of very active star formation on the west rim of \hii SGS LMC-2. The \hi position-velocity plot shows very complex velocity structure. The WR star is a member of LH103.

{\bf WR146} is on the northeast rim of the 7\farcm5 $\times$ 6\farcm3 superbubble in DEM\,L269 and on the western rim of \hii SGS LMC-2. The WR star is a member of LH104 which is responsible for the superbubble.

{\bf WR147} is outside the superbubble of DEM\,L269 in the ridge of active star formation on the western base of \hii SGS LMC-2. The WR star is $\sim$60 pc from LH104, and is thus likely not a member of this OB association.

{\bf WR148} is to the west of a long north-south oriented filament, out of which a faint 2\farcm8 $\times$ 1\farcm8 arc extends around the WR star, forming a shell structure. 
The bright long filament actually connects fainter filaments on both ends to form a large 6\farcm3 $\times$ 4\farcm5 shell.  The star is located in \hii SGS LMC-2, DEM\,L310, and to the east of 30 Dor. The \hi position-velocity plots show multiple velocity components, indicating complex kinematic structures, rendering the aforementioned apparent shells somewhat uncertain.

{\bf WR149} is near the small diffuse \hii region DEM\,L294.

{\bf WR150} is projected in the central cavity of \hii SGS LMC-2, or DEM\,L310. There are some filamentary features near the star, but nothing can unambiguously be identified to be associated with the star. 

{\bf WR151} is at the southern edge of the \hii region DEM\,L309 and projected in the northern interior of a large ionized shell measuring 17$'$ $\times$ 15$'$, which is associated with \hi GS\,94. The WR star is a member of LH116.

{\bf WR152} is projected on the northern exterior of the \hii region DEM\,L309 and is $\sim$55 pc from LH116.  The WR star is near the northern boundary of the large 
filamentary ionized shell measuring 17$'$ $\times$ 15$'$, which is associated with \hi GS\,94.

{\bf WR153} is superposed on faint diffuse emission of DEM\,L308 inside a large filamentary ionized shell measuring 17$'$ $\times$ 15$'$, which is associated with \hi GS\,94.  The \hi position-velocity plots show velocity splits indicating 
an expansion velocity of $\sim$20 \kms for \hi GS\,94.  The WR star is over 60 pc from LH116 and thus not likely a member of this OB association.

{\bf WR154} is projected in a small 1\farcm8 $\times$ 1\farcm3 elliptical shell in the western interior of a larger 7\farcm5 $\times$ 5\farcm2 shell in DEM\,L315 \citep{Chu1980}.  The expansion velocity of the small shell has been observed to be $\sim$47 \kms \citep{Chu1983}. DEM\,L315 is on the northeastern rim of a large 17$'$ $\times$ 15$'$ ionized shell associated with \hi GS\,94.



\end{CJK}
\end{document}